\documentclass[review]{elsarticle}

\usepackage{lineno,hyperref}
\modulolinenumbers[5]

\journal{arXiv}

\usepackage[margin=2.5cm]{geometry}% by courtesy of Mico
\usepackage[T1]{fontenc}    % use 8-bit T1 fonts
\usepackage{lmodern}
\usepackage{hyperref}       % hyperlinks
\usepackage{url}            % simple URL typesetting
\usepackage{booktabs}       % professional-quality tables
\usepackage{amsfonts}       % blackboard math symbols
\usepackage{nicefrac}       % compact symbols for 1/2, etc.
\usepackage{microtype}      % microtypography
\usepackage{lipsum}
\usepackage{graphicx}
\usepackage{framed,multirow}
\usepackage{amssymb}
\usepackage{ mathrsfs }

\usepackage{latexsym}
\usepackage{bm}

\usepackage{url}
\usepackage{xcolor}
\usepackage[utf8x]{inputenc}
\DeclareUnicodeCharacter{"2068}{}
\DeclareUnicodeCharacter{"2069}{}
\usepackage{amsmath}
\usepackage{amsthm}
\graphicspath{{images/}}
\usepackage{parskip}
\usepackage{fancyhdr}
\usepackage{xcolor}
\usepackage{pgf}
\usepackage[font=small,labelfont=bf]{caption}
\usepackage{subcaption}
\usepackage{epstopdf}
\usepackage{pgfplots}
\usepackage{color}
\usepackage{tikz}
\usetikzlibrary{shapes.geometric}
\usepackage{float}
\usepackage{bm}
\usepackage{stmaryrd}
\usepackage{cleveref}

\DeclareMathOperator*{\argmin}{argmin} 
\usetikzlibrary{arrows, arrows.meta}
\definecolor{newcolor}{rgb}{.8,.349,.1}
\usetikzlibrary{calc}
\usetikzlibrary{external}
\usepackage{calrsfs}
\pgfplotsset{compat=1.15} 
\usepackage{scrextend}
\newtheoremstyle{problemstyle}  % <name>
{3pt}                                               % <space above>
{3pt}                                               % <space below>
{\normalfont}                               % <body font>
{}                                                  % <indent amount}
{\bfseries}                 % <theorem head font>
{\normalfont\bfseries:}         % <punctuation after theorem head>
{.5em}                                          % <space after theorem head>
{}                                                  % <theorem head spec (can be left empty, meaning `normal')>
\theoremstyle{problemstyle}

\newtheorem{problem}{Problem}[section]

\newtheoremstyle{Assumptionstyle}  % <name>
{3pt}                                               % <space above>
{3pt}                                               % <space below>
{\normalfont}                               % <body font>
{}                                                  % <indent amount}
{\bfseries}                 % <theorem head font>
{\normalfont\bfseries:}         % <punctuation after theorem head>
{.5em}                                          % <space after theorem head>
{}                                                  % <theorem head spec (can be left empty, meaning `normal')>
\theoremstyle{Assumptionstyle}

\newtheorem{assumption}{Assumption}[section]

\newcommand{\vertiii}[1]{{\left\vert\kern-0.25ex\left\vert\kern-0.25ex\left\vert #1 
    \right\vert\kern-0.25ex\right\vert\kern-0.25ex\right\vert}}

\begin{document}

\begin{frontmatter}

\title{A Continuous $hp-$Mesh Model for Discontinuous Petrov-Galerkin Finite Element Schemes with Optimal Test Functions}

\author[1]{Ankit Chakraborty}
\cortext[mycorrespondingauthor]{Corresponding author}
\ead{chakraborty@aices.rwth-aachen.de}
%\ead{chakraborty@aices.rwth-aachen.de}
\address[1]{AICES, RWTH Aachen University, Schinkelstra\ss e 2, 52062 Aachen, Germany}

\author[2]{Georg May}

\address[2]{Von Karman Institute for Fluid Dynamics, Waterloosesteenweg 72, 1640 Sint-Genesius-Rode, Belgium}

\begin{abstract}
We present an anisotropic $hp-$mesh adaptation strategy using a continuous mesh model for discontinuous Petrov-Galerkin (DPG) finite element schemes with optimal test functions, extending our previous work \cite{CHAKRABORTY20221} on $h-$adaptation. The proposed strategy utilizes the inbuilt residual-based error estimator of the DPG discretization to compute both the polynomial distribution and the anisotropy of the mesh elements. In order to predict the optimal order of approximation, we solve local problems on  element patches, thus making these computations highly parallelizable. The continuous mesh model is formulated either with respect to the error in the solution, measured in a suitable norm, or with respect to certain admissible target functionals. We demonstrate the performance of the proposed strategy using several numerical examples on triangular grids. 
\end{abstract}


\end{frontmatter}

\linenumbers
\section{Introduction}
Automatic mesh adaptation algorithms are potent tools that aid the computation of efficient and accurate solutions of partial differential equations (PDEs). These algorithms can significantly increase the accuracy and computational efficiency by modifying the functional space in which the approximate solution is sought. These modifications can be roughly classified into three different categories:
\begin{itemize}
\item Modifying the geometrical properties of mesh elements ($h-$adaptation);
\item modifying the order of approximation space ($p-$adaptation);
\item combining $h-$ and $p-$adaptation ($hp-$adaptation).
\end{itemize}
In this article, we focus on $hp-$adaptation. An automatic $hp-$mesh adaptation strategy attempts to produce a near-optimal distribution of element size ($h-$adaptivity) and combine it with an appropriate distribution of local order of approximation ($p-$adativity) with the least amount of user's input. Such an automatic mesh adaptation strategy faces the problem of locally choosing between $h-$refinement or $p-$refinement. This issue has been addressed in the past in many isotropic mesh-adaptation methods~\cite{houston_adap,giani_adap,houston_adap2,jmhp,BABUSKA19905,LD_book}.

In recent times, metric-based mesh adaptation has emerged as a promising technology for generating meshes, especially for computational fluid dynamics \cite{ceze2013, Leicht2008, LEICHT20107344}. Meshes comprised of simplices can be represented by a Riemannian metric field. In metric-based mesh adaptation techniques, the parameters defining the metric field are optimized to generate an appropriate mesh. In general, this optimization problem is discrete and defies any analytical solution. However, using the continuous mesh approach \cite{inria_a,inria_b}, one formulates a continuous analog of the discrete mesh optimization problem, thus allowing analytical techniques, such as calculus of variations, to be employed for optimization. Although the mesh generated from the continuous mesh model is not provably optimal, it typically has excellent approximation properties \cite{Dolejsi2017,Rangarajan2018,CHAKRABORTY20221,RANGARAJAN2020109321,Ar2021}. When these adaptive techniques are combined with higher-order approximation methods, they present themselves as very potent tools for attaining higher accuracy with reduced degrees of freedom~\cite{cfd2030}. The potency of metric-based adaptation methods stems from the flexibility they provide for the efficient discretization of the computational domain near difficulties such as singularities, interior, and boundary layers. Consult \cite{ringue_adap,RANGARAJAN2020109321,Dolejsi2018,Dahm_dissert} for recent development in the context of metric-based adaptation for hybridized discontinuous Galerkin and discontinuous Galerkin discretizations.

An efficient mesh adaptation algorithm is only a part of the machinery required for computing accurate numerical solutions of PDEs. It complements the numerical method employed to compute the approximate solution. Typically, automatic mesh adaptation algorithms employ {some a posteriori} measure of error. These measures are computed using the approximate solution. Thus, the accuracy and stability of the underlying numerical scheme become indispensable for accurate mesh adaptations. In terms of robust and stable higher-order methods, discontinuous Petrov-Galerkin schemes (DPG) with optimal test functions have been a critical development over the past decade. Demkowicz and Gopalakrishnan first introduced the DPG methodology with optimal test functions in \cite{Demkowicz2010,dem_part2}. The core idea of this approach is to compute a discrete test space such that the best possible discrete inf-sup constant is achieved, thus ensuring higher numerical stability and accuracy. Apart from numerical accuracy and stability, these numerical schemes are accompanied by a residual-based error estimator. Due to the presence of this inbuilt error estimator, the DPG methodology with optimal test functions is tailor-made for mesh optimization \cite{Demkowicz2012a, KeithGoal, DEMKOWICZ2020}.
%such that the optimal test space can be complemented with an optimal approximation space. 

%The inbuilt error-estimate can be approximated using polynomial basis functions, thus giving it an anisotropic nature. The interplay between the anisotropy of the inbuilt error-estimator and the anisotropy of the mesh elements has motivated the work presented in this article.

This article proposes a metric-based anisotropic adaptation framework that uses the DPG inbuilt error estimator to generate meshes with optimized size and shape distribution, as well as an optimized polynomial degree distribution, thus complementing the optimal test space with a near-optimal approximation space. The article is structured as follows. First, section 2 briefly discusses PG schemes with optimal test functions. Then, in section 3, we review the relationship between a metric field and a  triangular mesh. The proposed method to optimize the local mesh anisotropy and the continuous $hp-$mesh model used to determine the optimal element size and polynomial distribution are discussed in section 4 and section 5, respectively. Finally, in section 6, we provide numerical results, and concluding remarks in section 7. 
\section{Discretization}
%\gm{should we have a subsection here, since we have another subsection below?}

This section briefly discusses the formulation of DPG schemes with optimal test functions. These schemes can be interpreted as minimum residual methods, which allows derivation of an inbuilt error-estimator in a straightforward way. Let $\mathbb{X}$ and $\mathbb{V}$ be Hilbert spaces over $\mathbb{R}^{d}$. We consider a well-posed variational boundary value problem using a continuous bilinear form $b:\mathbb{X} \times \mathbb{V} \rightarrow \mathbb{R}$. Here $\mathbb{X}$ is the approximation space, and $\mathbb{V}$ is the test space. Let $\mathbb{X}'$ and $\mathbb{V}'$ be the respective dual spaces of $\mathbb{X}$ and $\mathbb{V}$. For a given functional $\ell \in \mathbb{V}'$, the primal solution can be defined as $u^{\star} \in \mathbb{X}$ satisfying  
\begin{align}
b(u^{\star},v) = \ell(v) \quad \forall \quad v\in \mathbb{V}. \label{blnr_a}
\end{align}
The bilinear form $b$ generates a continuous linear operator $\mathcal{B}: \mathbb{X} \rightarrow \mathbb{V}'$ such that,
\begin{align}
\left\langle \mathcal{B}u,v \right\rangle = \left\langle u,\mathcal{B}'v \right\rangle = b(u,v) \quad {\forall} \quad u \in \mathbb{X},v\in\mathbb{V}.
\end{align}

 Thus, $u^{\star} = {\mathcal{B}}^{-1}\ell$. Let $\mathbb{X}_h \subset \mathbb{X}$ be a finite dimensional subspace of $\mathbb{X}$.  One can characterize the optimal solution $u_h^{opt} \in \mathbb{X}_h$ as the one with minimal error in an appropriately chosen norm. Since $u^{\star}$ is not accessible a priori, it cannot be invoked to define the optimality statement. Instead, the following minimum residual principle can be used
\begin{align}
u_h^{opt} = \argmin_{u_h \in \mathbb{X}_h} {\Vert \mathcal{B} u_h - \ell \Vert}^2_{\mathbb{V}'}.
\label{ip1}
\end{align} 
%defined by duality as $\langle \mathcal{R}_\mathbb{V} v, v^{\prime} \rangle = {\left( v, v^{\prime} \right)}_{\mathbb{V}} \, \forall \, v^{\prime} \in \mathbb{V}$. It can be observed that 
%\begin{align}
% {\Vert \mathcal{B} u_h - \ell \Vert}^2_{V'} =  \left\langle \mathcal{B} u_h - \ell, {\mathcal{R}_\mathbb{V}}^{-1} \left(\mathcal{B} u_h - \ell \right) \right\rangle \label{ip1}.
%\end{align}
On taking the Gateaux derivative of ${\Vert \mathcal{B} u_h - \ell \Vert}^2_{\mathbb{V}'}$ and using the first order optimality condition, the following variational equation is obtained:
\begin{align}
\left\langle \mathcal{B}u_h^{opt},{\mathcal{R}_\mathbb{V}}^{-1}\mathcal{B}u_h\right\rangle = \left\langle \ell, {\mathcal{R}_\mathbb{V}}^{-1}\mathcal{B}u_h\right\rangle \quad \forall u_h \in \mathbb{X}_h \label{optblnr_a},
\end{align}
where ${\mathcal{R}}_\mathbb{V}: \mathbb{V} \rightarrow {\mathbb{V}}^{\prime}$ is the Riesz map.
%The mapping $F_{opt} = {\mathcal{R}_\mathbb{V}}^{-1}\mathcal{B}:\mathbb{X} \rightarrow \mathbb{V}$ represents the mapping from the approximation space to the optimal test space. 
~\Cref{optblnr_a} can be redefined as
\begin{align}
b(u^{opt}_h,v_h) = \ell(v_h) \quad \forall v_h \in {\mathcal{R}_\mathbb{V}}^{-1}\mathcal{B}(\mathbb{X}_h).
\end{align}
On defining the energy inner product on $\mathbb{X}$ as $a(u,w) = \left\langle \mathcal{B}u, {\mathcal{R}_{\mathbb{V}}}^{-1}\mathcal{B}w \right\rangle$, we can define a norm on $\mathbb{X}$ as follows:
\begin{align}
\vertiii{u}_{\mathbb{X}} = a(u,u) = {\Vert \mathcal{B}u \Vert}^2_{\mathbb{V}'}.
\end{align}
%\begin{align}
%a(u,w) = \left\langle \mathcal{B}u, {\mathcal{R}_{\mathbb{V}}}^{-1}\mathcal{B}w \right\rangle,
%\end{align} 
%we can define a norm on $\mathbb{X}$ as follows:
%\begin{align}
%\vertiii{u}_{\mathbb{X}} = a(u,u) = {\Vert \mathcal{B}u \Vert}^2_{\mathbb{X}}.
%\end{align}

Since $\mathbb{V}$ is infinite-dimensional, the inverse of the Riesz map cannot be computed exactly. In practice, the Riesz map is approximately inverted over an enriched finite dimensional subspace $\mathbb{V}_r \subset \mathbb{V}$ such that $M = dim(\mathbb{V}_r) \geq  dim(\mathbb{X}_h) = N$ \footnote{In practice, the basis for $\mathbb{V}_r$ is enriched by adding higher-order polynomials to the basis of $\mathbb{X}_h$. Hence, if the basis functions of $\mathbb{X}_h$ are obtained using order $p$ polynomials, then we obtain the basis for $\mathbb{V}_r$ by employing polynomials of order $p + \delta p$. \label{fn_enrch}} \cite{VaziriAstaneh2018}. The approximation involves a symmetric Gram matrix which is induced by the inner product on $\mathbb{V}$. Thus, a practical implementation of the DPG method aims to find $u_h \in \mathbb{X}_h$ such that
%\begin{align}
%b(u_h,F_r w_h) = \ell(F_r w_h) \quad \forall  w_h \in \mathbb{X}_h, \label{apprx_blnr}
%\end{align}
\begin{align}
b(u_h,v_h) = \ell(v_h) \quad \forall  v_h \in \mathbb{V}_h.\label{apprx_blnr2}
\end{align}
where $\mathbb{V}_h = Range(F_r)$, and $F_r:\mathbb{X}_h \rightarrow \mathbb{V}_r$ represents the approximation of the mapping ${\mathcal{R}_\mathbb{V}}^{-1}\mathcal{B}$. 
%In particular, $F_r$ satisfies 
%\begin{align}
%(F_r w_h, v_r) = b(w_h,v_r) \quad \forall  w_h \in \mathbb{X}_h,v_r \in \mathbb{V}_r.
%\end{align}
%Since $dim(\mathbb{V}_r) \geq  dim(\mathbb{X}_h)$, $Range(F_r) \subset \mathbb{V}_r$. Let . Then~\cref{apprx_blnr} can be stated as:

Let  $\varphi$ be the Riesz representation of the residual, i.e. $\mathcal{R}_\mathbb{V} \varphi = (\mathcal{B}u_h - \ell) $. 
%\begin{align}
%%\varphi = {\mathcal{R}_\mathbb{V}}^{-1}(\mathcal{B}u_h - \ell) \\
%\mathcal{R}_\mathbb{V} \varphi = (\mathcal{B}u_h - \ell) , \label{oprt_err}
%\end{align}
Since it is not possible to compute the exact inverse of $\mathcal{R}_{\mathbb{V}}$, we need to approximate $\varphi$ using the finite-dimensional subspace $\mathbb{V}_r$, i.e., %Then, for , we have
\begin{align}
\left\langle \mathcal{R}_\mathbb{V} \varphi_h, \psi_i \right\rangle = {(\varphi_h,\psi_i)}_{\mathbb{V}} = \left\langle \mathcal{B}u_h - \ell, \psi_i \right\rangle \quad \forall \qquad i = 1,\dots M, \label{oprt_errb}
\end{align}
where $\psi_1, \dots ,\psi_M $  is any basis for $\mathbb{V}_r$ and {$\varphi_h$ approximates $\varphi$}. Let $\varphi_h = \sum_{j=1}^{M} \hat{c}_j \psi_j$. Then from~\cref{oprt_errb}, we have
\begin{align*}
\mathbb{G}{\bm{c}} = \bm{r}, 
\end{align*}
where $\mathbb{G} \in \mathbb{R}^{M \times M}$ is the Gram matrix with $\mathbb{G}_{i,j} = (\psi_i,\psi_j)_{\mathbb{V}}$, $\bm{c} = (\hat{c}_1, \dots, \hat{c}_M)^T$, and $\bm{r} \in \mathbb{R}^M$ with ${r}_i = b(u_h,\psi_i) - \ell(\psi_i)$. Thus, the error in energy norm is approximated by:
\begin{align}
\vertiii{ u - u_h }_{\mathbb{X}} = {\Vert \ell - \mathcal{B}u_h \Vert}_{\mathbb{V}'} \approx {\Vert \varphi_h \Vert}_\mathbb{V}. \label{err_est}
\end{align}
The function $\varphi_h$ is also known as the error-representation function \cite{Demkowicz2012a,Zitelli2011}. In this article, we employ the ultra-weak variational formulation for our numerical experiments. Details about the algebraic structure of the corresponding linear system can be found in \cite{VaziriAstaneh2018}. {The ultra-weak formulation for convection-diffusion and diffusion problems leads to a composite trial space i.e. $U_h:=(u_h,\sigma_h,\hat{u}_h,\hat{\sigma}_h) \in \mathbb{X}_h$ \footnote{{$\mathbb{X}_h \subset \mathbb{X} = L^2 \times {(L^2)}^d \times H^{\frac{1}{2}} \times H^{-1}$}} where $u_h$ approximates $u$, $\sigma_h$ approximates $\nabla u$, $\hat{u}_h$ approximates the trace of $u$ on the mesh skeleton, and $\hat{\sigma}_h$ approximates the normal flux on the mesh skeleton {with $u$ being the diffused or the convected quantity}. When presenting numerical results in~\cref{results_hp}, we refer to the approximate energy norm error ${\Vert \varphi_h \Vert}_{\mathbb{V}}$ as ${\Vert U - U_h \Vert}_E$.}

% Using a similar  process for the mapping $F_r$ as previously shown for $\psi_h$, one obtains the following linear system for the primal problem:
%\begin{align}
%B^T\mathbb{G}^{-1}B \hat{\mathbf{x}} = B^T\mathbb{G}^{-1}L.
%\end{align}
%Here $u_h = \sum_{i = 1}^N \hat{\mathbf{x}}_i \Phi_i$ with $\Phi_i$ being the basis functions of $\mathbb{X}_h$ and $\hat{\mathbf{x}} \in \mathbb{R}^N$. $B\in \mathbb{R}^{M \times N}$ is the enriched stiffness matrix where $B_{i,j} = b(\Phi_j,\psi_i)$ and $L$ is the enriched load vector with $L_i = \left\langle \ell, \psi_i \right\rangle$. In this work, we have employed the ultra-weak formulation with broken Sobolev spaces. This results in a block diagonal structure of $\mathbb{G}$ which allows us to invert $\mathbb{G}$ locally for each element. In our abstract setting we have only represented the error representation function with one component. For convection-diffusion and diffusion problems with ultra-weak formulations, the error-representation function has two components \cite{dem_part2, Demkowicz2012a, VaziriAstaneh2018}. Thus, we represent the error-representation function for convection-diffusion and diffusion problems by $(\varphi_v,\varphi_{\mathbf{\tau}})$ corresponding to two components of the test space $(v,\mathbf{\tau}) \in \mathbb{V}_r$. \gm{This section is very long. Can it be shortened?}

%\subsection{DPG-star method for the Dual problem}
Next, we briefly discuss the DPG-star method \cite{DEMKOWICZ2018}. We employ the DPG-star method to solve a compatible dual problem to obtain an error indicator that can drive goal-oriented adaptations. The DPG-star method has been previously used in this context, both for  isotropic $h-$adaptation \cite{KeithGoal} and anisotropic $h-$adaptation~\cite{CHAKRABORTY20221}. Let the strong form of the primal problem be given by
\begin{align}
Lu = s \quad \mathrm{in}\ \Omega, \quad Au = g \quad \mathrm{on} \ \partial \Omega,
\end{align}
where $\Omega$ is the computational domain, $L$ is a linear differential operator, $s \in L^2(\Omega)$, $g \in L^2(\partial \Omega)$, and $A$ is a linear differential boundary operator on $\partial \Omega$. In this article, we deal with target functionals of  the following form:
\begin{align}
J(u) = \int_{\Omega} j_{\Omega} u \, dx  + \int_{\partial \Omega} j_{\partial \Omega} Cu \, dx, \label{tf}
\end{align}
where  $j_{\Omega} \in L^2(\Omega)$, $j_{\partial \Omega} \in L^2(\partial \Omega)$, and $C$ is a boundary operator on $\partial \Omega$. We assume that the target functional satisfies the compatibility condition for linear problems \cite{Adjcompat}, i.e.,
\begin{equation}
{\left( {L}u,z \right)}_{\Omega} + {\left( Au,C^{\star}z \right)}_{\partial \Omega} = {\left( u, {{L}}^{\star}z \right)}_{\Omega} + {\left( Cu,A^{\star}z \right)}_{\partial \Omega},
\end{equation}
where $L^{\star},A^{\star}$ and $C^{\star}$ are the adjoint operators of $L,A$ and $C$. The DPG-star method approximates the solution of the dual problem associated with the target functional in~\cref{tf}.
The associated dual problem is given by
\begin{align}
{L}^{\star} z  = j_{\Omega} \quad \mathrm{in}\  \Omega, \quad
A^{\star}z   = j_{\partial \Omega} \quad \mathrm{on}  \partial \Omega.
\end{align}
An in-depth exposition to the DPG-star method can be found in \cite{DEMKOWICZ2018,KeithGoal}. The resulting error in the target functional can be bounded as:
\begin{align}
\vert J(u- u_h) \vert \leq  {\Vert \mathcal{B} \Vert}^{-1} {\Vert \mathcal{B} u_h - l \Vert}_{\mathbb{V}'} {\Vert {\mathcal{B}'} z_h - J \Vert}_{U'} .\label{err_bnd_impl}
\end{align} 
In~\cref{err_bnd_impl}, ${\Vert \mathcal{B} u_h - l \Vert}_{\mathbb{V}'}$ is the well-studied energy error estimator for DPG methods whereas for the residual of the dual problem, several error estimators have been proposed~\cite{DEMKOWICZ2018}. In the present work, we use the DPG-star element error indicator proposed in section 5.2 of \cite{DEMKOWICZ2018} for convection-diffusion problems with ultra-weak formulation:

\begin{equation}
\eta^{\star}_k =  {\left({\Vert {{L}}^{\star} (v_{h,z},\bm{\tau}_{h,z}) - j_{\Omega} \Vert }_{L^2(k)}^2 + \sum_{e \in \partial k \setminus \partial \Omega} h_e {\Vert {\llbracket \bm{\tau}_{h,z} \cdot \bm{n} \rrbracket} \Vert}_{L^2(e)}^2 + \sum_{e \in \partial k} {h_e}^{-1}  {\Vert {\llbracket v_{h,z} \rrbracket} \Vert}_{L^2(e)}^2 \right)}^{\frac{1}{2}}, \label{ele_dual_ind}
\end{equation}

where $v_{h,z},\bm{\tau}_{h,z}$ represent the approximate adjoint solution, $\llbracket \bm{\tau}_{h,z} \cdot \bm{n} \rrbracket$ and ${\llbracket v_{h,z} \rrbracket}$  represents the jump across the edge $e$ of length $h_e$. We assume that the variational problem is well-posed and $\mathcal{B}$ is bounded from below. Thus, we can use the product of the energy error estimator and the DPG-star error indicator as a valid error indicator for the goal-oriented mesh adaptations. 
Since we are using the DPG-star method to solve the dual problem, we have replaced ${L}$ with ${L}^{\star}$ in the error estimator from section 5.2 of \cite{DEMKOWICZ2018} to obtain~\cref{ele_dual_ind}.

% In~\cref{ele_dual_ind}, there are two dual solution components represented by $(v_{h,z},\bm{\tau}_{h,z})$, as the DPG-star method approximates the dual solution using an approximate optimal test space.  \gm{?} In order to compute the dual weighted residual (DWR) error estimator for correction of the target functional, we need to project the solution to $\mathbb{V}_r$ to circumvent the Galerkin orthogonality.
\section{Mesh representation using metric fields}
Next, we provide a short description of mesh representation using Riemannian metric fields. Such metric-based mesh representation provides a convenient way of encoding and manipulating a mesh~\cite{inria_a,inria_b}. In this work, we concern ourselves with meshes consisting of triangles, but the proposed ideas and formulations can be extended to tetrahedra \cite{Archk}. 
%Typically, meshes are considered to represent a large domain using smaller discrete elements. This discrete interpretation of a mesh leads to increased complexity and reduced flexibility in terms of formulating a data structure for encoding geometrical information of mesh elements. Thus, giving rise to the need for a mesh representation that allows a flexible encoding and manipulation of information pertaining to geometrical features of elements. 

\begin{figure}[h]
\begin{center}
\includegraphics[scale=1.0]{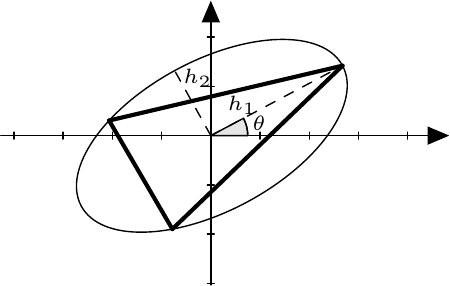}
\caption{Circumscribing ellipse and triangle for $C = 3$.}\label{ellip_ele}
\end{center}
\end{figure}

Let $T_h$ be a given triangulation. Then an element in $T_h$ can be characterized by a symmetric positive definite matrix. Let $k \in T_h$ represent an element with $e_i,\ i = 1,2,3$ being the vectorial representation of the edges. Then, there exists a non-degenerate symmetric positive definite matrix
\begin{align}
\mathbb{M} = \begin{bmatrix}
M_{1,1} & M_{1,2} \\
M_{2,1} & M_{2,2}
\end{bmatrix}
\end{align}
such that each edge is equilateral under the metric induced by $\mathbb{M}$, i.e.,
\begin{align*}
e_i^T \mathbb{M} e_i = C \quad \forall \quad i=1,2,3,
\end{align*}
where $C > 0$ is a constant. The triangular element can be associated with its circumscribing ellipse, given by $\{\mathbf{x}\in\mathbb{R}^2: \mathbf{x}^T\mathbb{M}\mathbf{x} = 1\}$.  This is illustrated in~\cref{ellip_ele}.

The eigenvectors and the eigenvalues of $\mathbb{M}$ encode information about the orientation ($\theta$) and the aspect ratio of the ellipse and, subsequently, the associated element. This is evident from the spectral decomposition of $\mathbb{M}$:
\begin{align}
\mathbb{M} =  {\begin{bmatrix}
\text{cos}(\theta) && -\text{sin}(\theta) \\
\text{sin}(\theta) && \text{cos}(\theta)
\end{bmatrix}}^T\begin{bmatrix}
\alpha_{1} && 0\\ 0 && \alpha_{2}
\end{bmatrix}\begin{bmatrix}
\text{cos}(\theta) && -\text{sin}(\theta) \\
\text{sin}(\theta) && \text{cos}(\theta)
\end{bmatrix}.
\end{align}

The eigenvalues $\alpha_1 = \frac{1}{h_1^2}$ and $\alpha_2 = \frac{1}{h_2^2}$ are related to the principal axes of the ellipse. The area of the triangle and the circumscribing ellipse are related by
\begin{align}
\vert k \vert  = \frac{3\sqrt{3}}{4} h_1 h_2 = \frac{3\sqrt{3}}{4d},
\end{align} 
where $d$ is called the local mesh density. Note that $d$ is proportional to inverse of area of the ellipse. The slenderness of the ellipse and the associated triangle is represented by the aspect ratio $\beta = \frac{h_1}{h_2} \geq 1$. With $\theta$, $\beta$, and $d$, one can describe the mesh element geometrically. Typically, most metric-based mesh adaptation algorithms generate a discrete metric field. Metric-based mesh generators suitably interpolate the discrete metric field to produce its continuous equivalent, which is further employed by the mesh generator to produce a metric conforming mesh \footnote{A metric conforming mesh contains elements which are nearly equilateral under the input metric field. In this article, we use BAMG\cite{Bamg} as the preferred metric-based mesh generator.}.

Next, we recall the concept of an $hp-$mesh. For a given triangulation $T_h$, we associate an integer $p_k$ with every element $k \in T_h$, representing its polynomial order of approximation. These integers form a vector called the polynomial distribution vector: $\mathbf{p} = (p_k)_{k \in T_h}$. The triangulation $T_h$ and the associated polynomial distribution vector $\mathbf{p}$ form the $hp-$mesh $T_{h,p} := \{T_h,\mathbf{p}\}$.
\section{Anisotropy computation}
In this section, we provide a brief description of the methodology employed for computing optimal anisotropy parameters $(\beta^{\star},\theta^{\star})$. In depth exposition can be found in our previous work \cite{CHAKRABORTY20221}. In order to compute the anisotropy parameters $(\beta^{\star},\theta^{\star})$, we employ the polynomial representation of the error estimator mentioned in~\cref{err_est}. In this article, we employ a scaled version of the standard inner product ($H^{1}(T_h) \times H(div,T_h)$) associated with the test space for convection-diffusion and diffusion problems. We call the corresponding test norm as the scaled V-norm~\cite{Demkowicz2012a,CHAKRABORTY20221}, given by
 \begin{align}
{\Vert (\psi_v,\bm{\psi}_{\tau}) \Vert}^2_{\mathbb{V},k} & = {\Vert \psi_{v} \Vert}^2_{2,k}  +  \sqrt{\vert k \vert} \,  {\Vert \nabla \psi_v \Vert}^2_{2,k}  + {\Vert \bm{\psi}_{\tau}\Vert}^2_{2,k}  +  \sqrt{\vert k \vert} \, {\Vert \nabla \cdot \bm{\psi}_{\tau}\Vert}^2_{2,k}  \notag \\ &= \int_k {(\psi_{v}(\mathbf{x}))}^2+ \bm{\psi_\tau}(\mathbf{x}) \cdot\bm{\psi}_{\tau}(\mathbf{x}) + \sqrt{\vert k \vert} ({\nabla \psi_{v}(\mathbf{x}) \cdot \nabla \psi_{v}(\mathbf{x})}  + {(\nabla \cdot\bm{\psi}_{\tau}(\mathbf{x}))}^2))\,d\mathbf{x}.  \label{scaled_norm_MN}
\end{align}
Here ${\vert k \vert}$ represents the volume of the element $k \, \in \, T_h$ and $(\psi_v,\bm{\psi}_{\tau})$ are the associated error-representation functions. The computation of optimal anisotropy parameters $(\beta^{\star},\theta^{\star})$ only involves minimization of a bound on the element's energy error estimate  (proposed in~\cite{CHAKRABORTY20221}, Section 4). This bound is given by a parameterized integral of $\beta$ and $\theta$. The minimum of the bound is sought  iteratively by performing alternate searches in $\beta$ and $\theta$~\cite{Dolejsi2019}.
\section{Continuous $hp-$mesh model} \label{hpadap}
In this section, we present the procedure for $hp-$adaptations using the inbuilt error estimator that accompanies DPG schemes with optimal test functions. The procedure is a two-step process that comprises selecting the order of polynomial approximation, which is followed by a mesh density computation. Next, we reiterate an important assumption (see \cite{CHAKRABORTY20221, Dolejsi2017}) which is fundamental for the proposed methodology:

\begin{assumption}
Let ${T}_h$ be a given triangulation and  $\eta_k$ be a local error estimate such that ${\eta}^2 = \sum_{k \in T_h} {\eta}^2_k$ and $\eta = O(h^s)$ i.e the error estimate converges with order $s$~\cite{Dolejsi2017}. We assume that the local error estimate $\eta_k$ scales\footnote{To motivate the scaling $\eta_k \propto {\vert k \vert}^{(s+1)}$, recall that the global error estimate $\eta^2$ scales as $h^{2s} \propto {\vert K \vert}^s$ (We have a method of order $s$.). At the same time, the contribution from each individual sub-element (obtained from local isotropic refinement) scales with an additional factor of k, because of the local domain of integration. For more details, see section 5, \cite{Dolejsi2017}} as $\eta_k = \overline{A}_k {\vert k \vert}^{(s+1)}$,  where $\overline{A}_k$ depends upon the anisotropy of the element and order of the method $s$, but is independent of $\vert k \vert$. Furthermore, the order of the method $s$ directly depends upon the polynomial degree of approximation $p$. \label{assump1}
\end{assumption}

Verification of~\cref{assump1} can be found in \cite{Dolejsi2017,VENDITTI200240}. Let $\left\{{T}_h\right\}_n$ be the sequence of triangulations employed. We intend to construct an error estimate which asymptotically achieves global equality with the inbuilt energy error estimate :
\begin{align}
\Vert U - U_h \Vert^2_{E,k} \approx e_d(\mathbf{x}_k) \vert k \vert \quad \text{for} \, \, \mathbf{x}_k \in k, \, k \in T_h, \label{origin}
\end{align}
where we define $e_d(\mathbf{x}_k)$ as the error density function. Thus  asymptotically ($h \rightarrow 0$), we have
\begin{align*}
\Vert U-U_h \Vert^2_{E,\Omega} = \sum_{k \in {T}_h} e_d(\mathbf{x}_k) \vert k \vert \rightarrow \int_{\Omega}  e_d(\mathbf{x}) \, d \mathbf{x}.
\end{align*}
Using~\cref{assump1}, $\vert k \vert = \frac{\alpha}{d}$, and~\cref{origin}, we define the error density function as follows:
 \begin{equation}
 e_d(\mathbf{x},p(\mathbf{x})) := \overline{A}(\mathbf{x},p(\mathbf{x})){d(\mathbf{x})}^{-s(\mathbf{x})}\alpha^{s(\mathbf{x})}, \label{error_density}
 \end{equation}
where $\alpha = \frac{3\sqrt{3}}{4}$. This asymptotic continuous-mesh error model will be used in our $hp$-mesh optimization scheme below. Using the a priori established order of convergence of the energy estimate~\cite{Demkowicz2011a}, we can set $ s(\mathbf{x}) = p(\mathbf{x})+1 $. In an $hp-$adaptive scheme, $s$ can vary for various elements in the mesh.
\subsection{Polynomial selection} \label{polyselec}
\begin{figure}[!h]
\begin{center}
\includegraphics[scale=0.7]{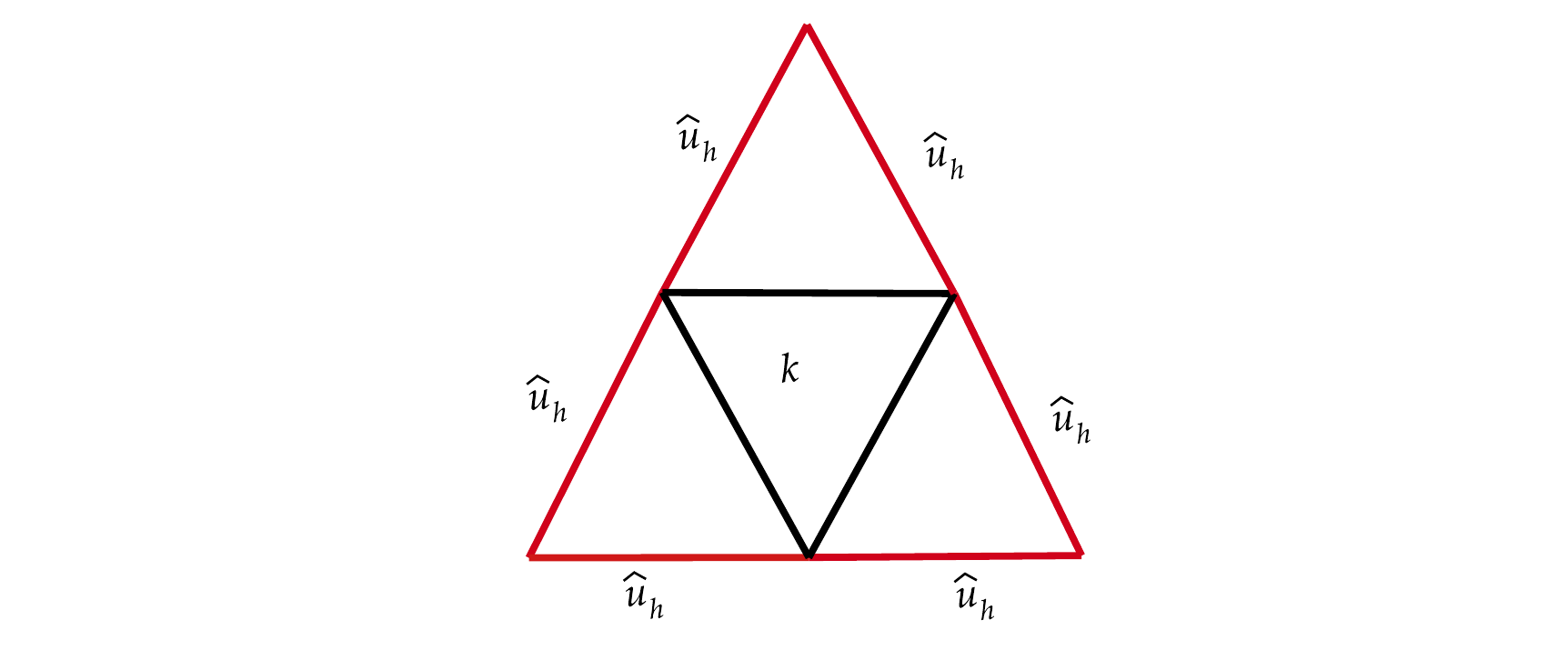}
\end{center}
\caption{Patch around an element $k$ over which local problems are solved. Edges marked by red represent Dirichlet boundaries of the patch.} \label{local_patch}
\end{figure}

To choose the order of polynomial approximation for an element, we solve the governing PDE locally over a patch surrounding the element. The boundary conditions for these local problems are obtained using the trace $(\hat{u}_h)$ for Dirichlet boundary conditions or the normal flux $(\hat{\sigma}_h)$ for Neumann boundary conditions (the traces or the normal fluxes utilized as the local boundary conditions are computed globally for the current polynomial distribution). For complex non-linear problems, choosing local boundary conditions is non-trivial and would need specific analysis. One such example would be the Euler equations, for which one can employ characteristic boundary conditions. In~\cref{local_patch}, we show an internal patch with Dirichlet boundary condition for a scalar convection-diffusion problem.    
These local problems are solved at three different polynomial orders, $p_k$\footnote{To have same fidellity of solution from the local solves at different polynomial orders, we solve the local problem at $p_k$ rather than using the solution obtained by solving the global system on the current $hp-$mesh.} and $p_k \pm 1$ where $p_k$ represents the current polynomial order in element $k \in T_h$. Once the solution is computed, it is followed by computing the local energy error estimate ${\Vert U- U_h \Vert}_{E,k}$ for each polynomial order. Let $E_{p_k+i}$ and $N_{p_k + i}$ denote, respectively, the energy error and the number of degrees of freedom, computed at the polynomial order $p_k+i$, where $i = -1,0,1$. Next, we introduce a parameter $m_{p_k + i}$ which corresponds to the amount of uniform refinement or coarsening required by $p_k + i$ to achieve the same error as $p_k$.
\begin{equation}
m_{p_k+i} = {\left(\frac{E_{p_k + i}}{E_{p_k}}\right)}^{\frac{2}{s_i +1}} N_{p_k + i}. \label{ref_par}
\end{equation}
In~\cref{ref_par}, $s_i$ is the a priori rate of convergence for order $p_k + i$. For the energy norm, we have used $s_i = p_k + i + 1$. In order to choose the optimal polynomial order, we compute $m_{p_k+i}$ for $p_k - 1$ and $p_k + 1$. The optimal order ($p_{k,opt}$) is the one which achieves $E_{p_k}$ with the smallest value of $m_{p_k+i}$, i.e.,
\begin{equation}
p_{k,opt} = \argmin_{i = -1,0,1} m_{p_k + i}.
\end{equation}
The rationale behind this approach is to find (if possible) a polynomial order, which is more efficient in terms of degrees of freedom required to achieve the same error level as $p_k$. For $i = 0$, the above computation is trivial. 

After we compute the polynomial distribution, we still need to compute the mesh density distribution. For this purpose, we target the continuous error estimate $E_c = \int_{\Omega} e_d(\mathbf{x}) \, d\mathbf{x}$ and solve a continuous minimization problem using calculus of variations. This is addressed in the next section.  The enrichment $\delta p$ %\footref{fn_enrch} 
of the space $\mathbb{V}_r$ when performing $hp-$adaptations is one order higher compared to the enrichment utilized when performing $h-$adaptation \cite{CHAKRABORTY20221}. This increase in order of enrichment stems from the use of approximation spaces of (up to) order $p_k + 1$ while solving the local problem.
% at  increases the required order of approximation for the $\mathbb{V}_r$. 
%and results in solution at $p_k + 1$ having the same fidelity as $p_k$ and $p_{k-1}$.

\subsection{Mesh density computation} \label{step_1}

In order to generate the optimal density distribution at fixed cost, we need to define the notion of cost. To that end, we define the mesh complexity  $\mathcal{N}_{h,p}$ as
\begin{align}
\mathcal{N}_{h,p} = \int_{\Omega} w(p(\mathbf{x})) d(\mathbf{x}) \, d\mathbf{x},
\end{align}
where
\begin{align}
w(p(\mathbf{x}) = \frac{(p(\mathbf{x})+1)(p(\mathbf{x})+1)}{2}.
\end{align}
The relation between the mesh complexity and number of degrees of freedoms $(N)$ can be computed as follows:
\begin{align}
\mathcal{N}_{h,p} = \int_{\Omega} w(\mathbf{x},p(\mathbf{x}))d(\mathbf{x})\, d\mathbf{x} \approx \sum_{k \in T_h} d(\mathbf{x}_k)w(\mathbf{x}_k,p_k) \vert k \vert =  \sum_{k \in T_h} \alpha w(\mathbf{x}_k,p_k) =  N(Ne,\mathbf{p}). \label{rel_cont_comp}
\end{align}
where $Ne$ is the number of elements and $\mathbf{p}$ is the polynomial distribution vector associated with the $hp-$mesh. Previously, we defined $\overline{A}(\mathbf{x})$ as a continuous analog of $\overline{A}_k$ from~\cref{assump1}. Similarly, we can treat the polynomial distribution vector $\mathbf{p}$ and introduce a continuous analog $p(\mathbf{x})$ for the continuous mesh model. 
%\gm{We have already used p(x) above (eq.(21)). Maybe move up?} 
The polynomial distribution vector $\mathbf{p}$ can be seen as a vector created from the snapshots of $p(\mathbf{x})$ in each element. The continuous analog of $\mathbf{p}$ and $\overline{A}_k$ allows us to formulate a continuous optimization problem as follows:

\begin{problem}
Let $\mathcal{N}_{h,p} $ be the desired mesh complexity and $e_d(\mathbf{x},p(\mathbf{x}))$ be the error density. We seek a mesh density distribution $d(\mathbf{x}): \Omega \rightarrow \mathbb{R}^{+}$ for a given polynomial distribution $p(\mathbf{x}): \Omega \rightarrow Z^{+}$ such that: \label{opt_prob}

(a) $\mathcal{N}_{h,p} = \int_{\Omega} w(\mathbf{x},p(\mathbf{x})) d(\mathbf{x}) \,d\mathbf{x }\qquad with \quad w(\mathbf{x},p(\mathbf{x})) = \frac{2(p(\mathbf{x})+1)(p(\mathbf{x})+2)}{3\sqrt{3}} $. 

(b) $E_c =\int_{\Omega} e_d(\mathbf{x},p(\mathbf{x})) \, d\mathbf{x}$ is minimized. 
\end{problem}
%For density computation, we have implemented the analytic optimization for $hp-$ continuous mesh model.
Using~\cref{error_density}, the continuous error estimate can be written as 
\begin{equation}
E_c = \int_{\Omega} {\alpha}^{(p(\mathbf{x})+1)}{\overline{A}(\mathbf{x},p(\mathbf{x}))} {d(\mathbf{x})}^{-(p(\mathbf{x})+1)} \, d\mathbf{x}. \label{contglobalerror}
\end{equation}

%In case of $hp-$adaptation, we use $s = p(\mathbf{x}) + 1$ as local order of convergence in~\cref{assump1}. 
In order to compute the minimum, we employ calculus of variations. Taking the variation of the complexity constraint in~\cref{opt_prob} with respect to $d(\mathbf{x})$ yields

\begin{equation}
\delta \mathcal{N}_{h,p}  = \int_{\Omega} w(\mathbf{x},p(\mathbf{x})) \delta d(\mathbf{x}) d\mathbf{x} = 0. \label{constraint_var}
\end{equation}

Next, on taking the variation of the continuous error estimate with respect to the density, we obtain
\begin{equation}
\delta E_c =   \int_{\Omega} -(p(\mathbf{x})+1) {\alpha}^{(p(\mathbf{x})+1)}{\overline{A}(\mathbf{x},p(\mathbf{x}))}{d}^{-(p(\mathbf{x})+2)} \delta d \, d\mathbf{x}.
\label{variation_energy}
\end{equation}

From~\cref{constraint_var} and~\cref{variation_energy}, it is implied that
\begin{equation}
\frac{{(p(\mathbf{x})+1)\overline{A}(\mathbf{x},p(\mathbf{x}))}}{w(\mathbf{x},p(\mathbf{x}))}  {\alpha}^{(p(\mathbf{x})+1)} {d(\mathbf{x})}^{-{(p(\mathbf{x})+2)}} = const. \label{comb_const}
\end{equation}
 
Solving~\cref{comb_const} for $d(\mathbf{x})$ yields

\begin{equation}
d^{\star}(\mathbf{x}) = {\left(  \frac{(p(\mathbf{x})+1)\overline{A}(\mathbf{x},p(\mathbf{x})) {\alpha}^{(p(\mathbf{x})+1)}}{w(\mathbf{x},p(\mathbf{x}))} \right)}^{\frac{1}{(p(\mathbf{x})+2)}} {const}^{-\frac{1}{(p(\mathbf{x})+2)}}. \label{opt_den_c}
\end{equation} 
 
Here, $const$ does not have a closed form solution. On substituting the expression for optimal density from~\cref{opt_den_c} into the  complexity constraint, we obtain
\begin{align}
\mathcal{N}_{h,p} = \int_{\Omega} w(\mathbf{x}){\left(  \frac{(p(\mathbf{x})+1)\overline{A}(\mathbf{x},p(\mathbf{x})) {\alpha}^{(p(\mathbf{x})+1)}}{w(\mathbf{x},p(\mathbf{x}))} \right)}^{\frac{1}{(p(\mathbf{x})+2)}} {const}^{-\frac{1}{(p(\mathbf{x})+2)}} \,  d\mathbf{x}. \label{bisec}
\end{align}

~\Cref{bisec} is non-linear in $const$ due to the varying exponent. Thus, we need to employ numerical techniques to solve for $const$. We use a bisection method to compute this constant in our current work. Once $const$ is computed, we can substitute its value  into~\cref{opt_den_c} to obtain the optimum density distribution $d^{\star}(\mathbf{x})$. 

The proposed $hp-$adaptive algorithm is a generalization of the $h-$adaptation  \cite{CHAKRABORTY20221}. If the variable polynomial distribution is substituted with $p(\mathbf{x}) = p$ throughout the domain, then $w(\mathbf{x})$ is constant and the proposed iterative algorithm reverts back to the $h-$only adaptation method. 
Since we are in a discrete setting in terms of triangulation, the quantities $\overline{A}(\mathbf{x},p(\mathbf{x}))$, $p(\mathbf{x})$ and optimal density $d^{\star}(\mathbf{x})$ are computed for each element $k \in T_h$. Thus, if $\mathbf{x}_k \in k$ and $k \in T_h$, using~\cref{assump1}, we have
\begin{align}
\overline{A}_{k,p_{k,opt}} = \frac{{\Vert U - U_h \Vert}^2_{E,k,p_{k,opt}}}{{\vert k \vert}^{p_{k,opt}+2}}  \label{abar_el}
\end{align}
and
\begin{equation}
d^{\star}(\mathbf{x}_k) = {\left(  \frac{(p_{k,opt}+1)\overline{A}_{k,p_{k,opt}} {\alpha}^{(p_{k,opt}+1)}}{w(\mathbf{x}_k,p_{k,opt})} \right)}^{\frac{1}{(p_{k,opt}+2)}} {const}^{-\frac{1}{(p_{k,opt}+2)}},  \label{opt_den_el}
\end{equation} 

where ${\Vert U - U_h \Vert}_{E,k,p_{k,opt}}$  represents the energy error in element $k$ for the optimal polynomial order ($p_{k,opt}$)  chosen via the process mentioned in~\cref{polyselec}.

\subsection{Goal-oriented adaptation}
For goal-oriented adaptation, we use the product of the DPG-star error estimate ($ \eta^{\star}_k$, see~\cref{ele_dual_ind}) and the energy estimate, i.e., $\eta_k =  \eta^{\star}_k {\Vert U - U_h \Vert}_{E,k,p_{k,opt}}$ in~\cref{assump1} to compute $\overline{A}_{k,p_{k,opt}}$. The polynomial order selection process stays the same, as we only depend upon the primal variables and their effect on energy error. For goal-oriented adaptations, we employ $s(\mathbf{x}) = p(\mathbf{x}) + 1$ in density computations. As previously mentioned in section 5.1 of \cite{CHAKRABORTY20221},~\cref{assump1} requires sufficient regularity but typically, the regularity of both dual and primal solution can not always be guaranteed. Hence, we pursue a pessimistic approach rather then setting $s(\mathbf{x}) = 2p(\mathbf{x}) + 1$ for convection-diffusion and diffusion problems.
\section{Numerical results}\label{results_hp}

\subsection{Test case I: boundary layer}\label{blhp}
Sharp boundary layers are one of the most encountered features in fluid dynamics. Through this test case, we present the fidelity of the proposed algorithm in the  presence of such boundary layers. In particular, we solve,
\begin{equation}
\begin{aligned}
\beta \cdot {\nabla}u-\epsilon{\nabla}^2u &= s \qquad&& \mathrm{in}\ \Omega = {(0,1)}^2, \\
u &= 0 && \mathrm{on} \ \partial \Omega, 
\end{aligned}
\end{equation}
where $\beta = {[1,1]}^T$ and $\epsilon > 0$. The source term $s(\mathbf{x})$ is chosen in such a manner that the exact solution is given by
\begin{equation}
u(\mathbf{x}) = \left( x + \frac{e^{\frac{x}{\epsilon}}-1}{1-e^{\frac{1}{\epsilon}}} \right) \left( y + \frac{e^{\frac{y}{\epsilon}}-1}{1-e^{\frac{1}{\epsilon}}} \right),
\end{equation}
%The solution exhibits a sharp boundary layer near $x \approx 1$, $y \approx 1$, increasingly so for smaller values of $\epsilon$. 
where $\mathbf{x} = [x,y]^T$. The strength of the boundary layer is inversely proportional to $\epsilon$. The $hp-$adaptation is initialized with a mesh comprised of $32$ elements and a constant polynomial order of $p_{initial} = 2$. Mesh complexity ($\mathcal{N}_{h,p}$) for the first adaptation cycle is computed as
\begin{equation}
\mathcal{N}_{h,p} = Ne \times \frac{\left(p_{initial}+1\right)\left(p_{initial}+2\right)}{2}\times \frac{3\sqrt{3}}{4},
\end{equation}
where $Ne$ represents the number of elements in the mesh. Between each adaptation cycle, $\mathcal{N}_{h,p}$ is increased by $30 \%$. The choice of growth in $\mathcal{N}_{h,p}$ is arbitrary. A different choice in growth may result in different pre-asymptotic behavior but should produce a similar asymptotic result. 
\begin{figure}[h!]
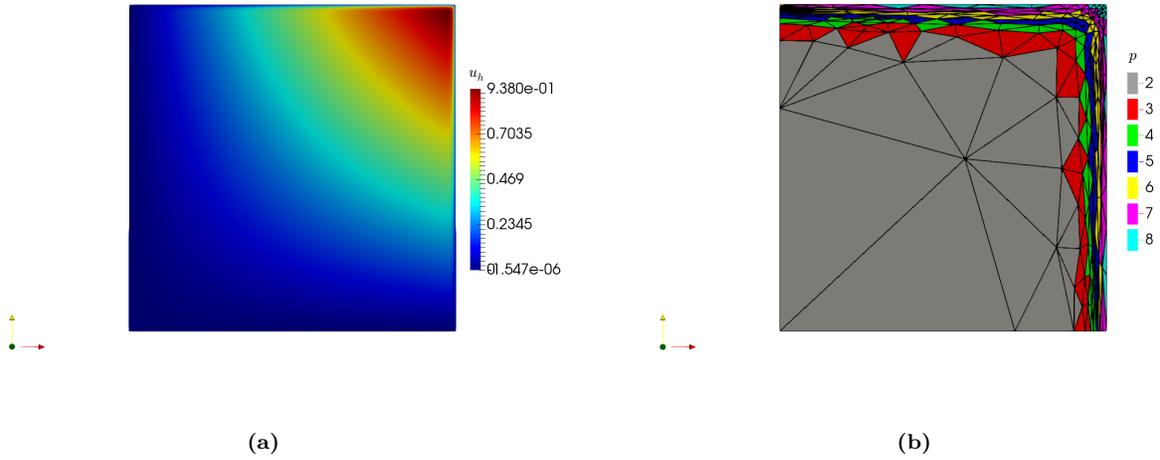

%\begin{center}
\begin{subfigure}[]{0.5\textwidth}
\includegraphics[scale=0.22]{Data_hp/Boundary_layer⁩/MNS_0p5/Boundarylayerhp_1to10_eps_0p005_296el_6980ndof-eps-converted-to.pdf}
\caption{}
\end{subfigure}
\hspace{0.2cm}
\begin{subfigure}[]{0.5\textwidth}
\includegraphics[scale=0.22]{Data_hp/Boundary_layer⁩/MNS_0p5/Boundarylayerhp_1to10_eps_0p005_296el_6980ndof_polyorder-eps-converted-to.pdf}
\caption{}
\end{subfigure}
\caption{Boundary layer: (a) solution contour on an adapted mesh and (b) polynomial distribution on the same adapted mesh with $\epsilon = 0.005$ and 6980 degrees of freedom. }
\label{fig:contourAndPdist}
%\end{center}
\end{figure}

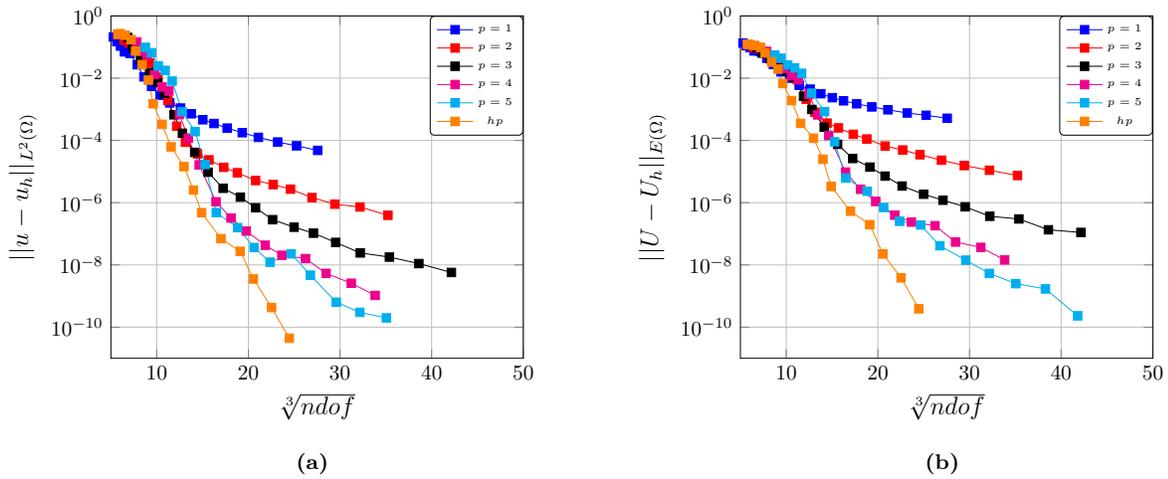
\begin{figure}[h!]
\begin{subfigure}[b]{0.5\textwidth}
\begin{tikzpicture}[scale=0.8]
		\begin{semilogyaxis}[xmin=5,xmax=50, ymin=1e-11,ymax=1,xlabel=\large{$\sqrt[3]{ndof}$},ylabel=\large{$||u-u_h||_{L^{2}(\Omega)}$},grid=major,legend style={at={(1,1)},anchor=north east,font=\tiny,rounded corners=2pt}]
		\addplot[color = blue,mark=square*] table[x= ndof, y=err_l2, col sep = comma] {Data_hp/Boundary_layer⁩/exp/L2_error_DPG_BL_0p005_p1_MN_0p5.txt};
		\addplot [color = red,mark=square*] table[x= ndof, y=err_l2, col sep = comma] {Data_hp/Boundary_layer⁩/exp/L2_error_DPG_BL_0p005_p2_MN_0p5.txt};
		\addplot [color = black,mark=square*] table[x= ndof, y=err_l2, col sep = comma] {Data_hp/Boundary_layer⁩/exp/L2_error_DPG_BL_0p005_p3_MN_0p5.txt};
		\addplot [color = magenta,mark=square*] table[x= ndof, y=err_l2, col sep = comma] {Data_hp/Boundary_layer⁩/exp/L2_error_DPG_BL_0p005_p4_MN_0p5.txt};
			\addplot [color = cyan,mark=square*] table[x= ndof, y=err_l2, col sep = comma] {Data_hp/Boundary_layer⁩/exp/L2_error_DPG_BL_0p005_p5_MN_0p5.txt};
%			\addplot [color = green,mark=square*] table[x= ndof, y=err_l2, col sep = comma] {Data_hp/Boundary_layer⁩/MNS_0p5/L2_error_DPG_BL_0p005_Hp_MN_0p5.txt};
           \addplot [color =orange,mark=square*] table[x= ndof, y=err_l2, col sep = comma] {Data_hp/Boundary_layer⁩/exp/L2_error_BL_MNS_0p5_Hpnew.txt};		
                   				
		\legend{$p = 1$,$p = 2$,$p = 3$,$p = 4$,$p = 5$,$hp$}
		\end{semilogyaxis}
	\end{tikzpicture}
	\caption{}
\end{subfigure}
\begin{subfigure}[b]{0.5\textwidth}
\begin{tikzpicture}[scale=0.8]
		\begin{semilogyaxis}[xmin=5,xmax=50, ymin=1e-11,ymax=1,xlabel=\large{$\sqrt[3]{ndof}$},ylabel=\large{$||U-U_h||_{E(\Omega)}$},grid=major,legend style={at={(1,1)},anchor=north east,font=\tiny,rounded corners=2pt}]
		\addplot[color = blue,mark=square*]  table[x= ndof, y=EE, col sep = comma] {Data_hp/Boundary_layer⁩/exp/EE_error_DPG_BL_0p005_p1_MN_0p5.txt};
		\addplot [color = red,mark=square*] table[x= ndof, y=EE, col sep = comma] {Data_hp/Boundary_layer⁩/exp/EE_error_DPG_BL_0p005_p2_MN_0p5.txt};
			\addplot [color = black,mark=square*] table[x= ndof, y=EE, col sep = comma] {Data_hp/Boundary_layer⁩/exp/EE_error_DPG_BL_0p005_p3_MN_0p5.txt};
					\addplot [color = magenta,mark=square*] table[x= ndof, y=EE, col sep = comma] {Data_hp/Boundary_layer⁩/exp/EE_error_DPG_BL_0p005_p4_MN_0p5.txt};
			\addplot [color = cyan,mark=square*] table[x= ndof, y=EE, col sep = comma] {Data_hp/Boundary_layer⁩/exp/EE_error_DPG_BL_0p005_p5_MN_0p5.txt};
			
\addplot [color =orange,mark=square*] table[x= ndof, y=EE, col sep = comma] {Data_hp/Boundary_layer⁩/exp/EE_error_BL_MNS_0p5_Hpnew.txt};

		\legend{$p = 1$,$p = 2$,$p = 3$,$p = 4$,$p = 5$,$hp$}
		\end{semilogyaxis}
\end{tikzpicture}
\caption{}
\end{subfigure}	
\caption{Convergence plots of (a) $L^2$ error in $u_h$ and (b) energy error using scaled V-norm.} \label{convergence_BL_scaled_math_norm_hp}
\end{figure}

\begin{figure}[h!]
\begin{subfigure}[b]{0.5\textwidth}
\begin{tikzpicture}[scale=0.8]
		\begin{axis}[xmin=0,xmax=18, ymin=1,ymax=600,xlabel=\large{$Adaptations$},ylabel=\large{$Ne$},grid=major,legend style={at={(1,1)},anchor=north east,font=\tiny,rounded corners=2pt}]
		\addplot [color = red,mark=square*] table[x= adap, y=ne, col sep = comma] {Data_hp/Boundary_layer⁩/MNS_0p5/order_ne_adap_p1to10_0p1.txt};	
		\addplot [color = blue,mark=square*] table[x= adap, y=ne, col sep = comma] {Data_hp/Boundary_layer⁩/MNS_0p5/order_ne_adap_p1to10_0p01.txt};	
		\addplot [color = black,mark=square*] table[x= adap, y=ne, col sep = comma] {Data_hp/Boundary_layer⁩/MNS_0p5/order_ne_adap_p1to10_0p005.txt};	
		\legend{$\epsilon = 0.1$,$\epsilon = 0.01$,$\epsilon = 0.005$}
		\end{axis}
	\end{tikzpicture}
	\caption{}
\end{subfigure}
\begin{subfigure}[b]{0.5\textwidth}
\begin{tikzpicture}[scale=0.8]
		\begin{axis}[xmin=0,xmax=18, ymin=1,ymax=15,xlabel=\large{$Adaptations$},ylabel=\large{$p_{avg}$},grid=major,legend style={at={(1,1)},anchor=north east,font=\tiny,rounded corners=2pt}]
		\addplot [color = red,mark=square*] table[x= adap, y=orderavg, col sep = comma] {Data_hp/Boundary_layer⁩/MNS_0p5/order_ne_adap_p1to10_0p1.txt};	
		\addplot [color =blue,mark=square*] table[x= adap, y=orderavg, col sep = comma] {Data_hp/Boundary_layer⁩/MNS_0p5/order_ne_adap_p1to10_0p01.txt};	
		\addplot [color = black,mark=square*] table[x= adap, y=orderavg, col sep = comma] {Data_hp/Boundary_layer⁩/MNS_0p5/order_ne_adap_p1to10_0p005.txt};	
		
		\legend{$\epsilon = 0.1$,$\epsilon = 0.01$,$\epsilon = 0.005$}
		\end{axis}
	\end{tikzpicture}
\caption{}
\end{subfigure}	
\caption{Evolution of (a) number of mesh elements and (b) average polynomial order with adaptations using the scaled V-norm at a fixed cost $\mathcal{N} = 3072$ for different values of $\epsilon$.} \label{hp_fixed_dof_ne_pavg}
\end{figure}
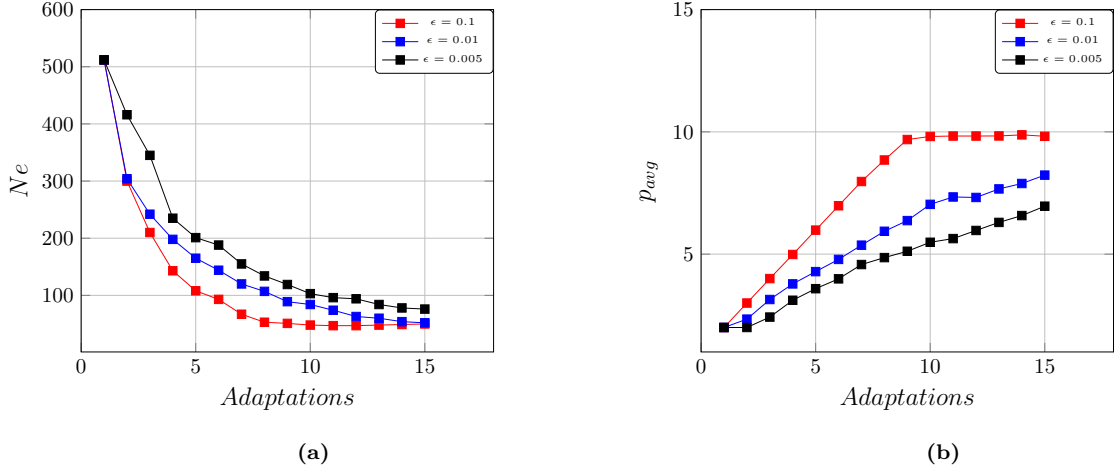
\begin{figure}[h!]
\begin{subfigure}[b]{0.5\textwidth}
\begin{tikzpicture}[scale=0.8]
		\begin{semilogyaxis}[xmin=5,xmax=30, ymin=1e-12,ymax=1,xlabel=\large{$\sqrt[3]{ndof}$},ylabel=\large{$||u-u_h||_{L^{2}(\Omega)}$},grid=major,legend style={at={(1,1)},anchor=north east,font=\tiny,rounded corners=2pt}]

           \addplot [color =blue,mark=square*] table[x= ndof, y=err_l2, col sep = comma] {Data_hp/Boundary_layer⁩/exp/L2_error_patchsolve_hp_1to20_DLS_128el_BL.txt};

           \addplot [color =red,mark=square*] table[x= ndof, y=err_l2, col sep = comma] {Data_hp/Boundary_layer⁩/exp/L2_error_patchsolve_hp_1to20_DLS_randompolyorder_BL.txt};	

		\legend{$p_{initial} = 2$, $p_{initial} = Random$}
		\end{semilogyaxis}
	\end{tikzpicture}
	\caption{}
\end{subfigure}
\begin{subfigure}[b]{0.5\textwidth}
\begin{tikzpicture}[scale=0.8]
		\begin{semilogyaxis}[xmin=5,xmax=30, ymin=1e-11,ymax=1,xlabel=\large{$\sqrt[3]{ndof}$},ylabel=\large{$||U-U_h||_{E(\Omega)}$},grid=major,legend style={at={(1,1)},anchor=north east,font=\tiny,rounded corners=2pt}]
		
\addplot [color =blue,mark=square*] table[x= ndof, y=EE, col sep = comma] {Data_hp/Boundary_layer⁩/exp/EE_error_patchsolve_hp_1to20_DLS_128el_BL.txt};

\addplot [color =red,mark=square*] table[x= ndof, y=EE, col sep = comma] {Data_hp/Boundary_layer⁩/exp/EE_error_patchsolve_hp_1to20_DLS_randompolyorder_BL.txt};

		\legend{$p_{initial} = 2$, $p_{initial} = Random$}
		\end{semilogyaxis}
\end{tikzpicture}
\caption{}
\end{subfigure}	
\caption{Convergence plots of (a) $L^2$ error in $u_h$ and (b) energy error using scaled V-norm with different initial polynomial distributions.} \label{convergence_BL_randompolyorder}
\end{figure}
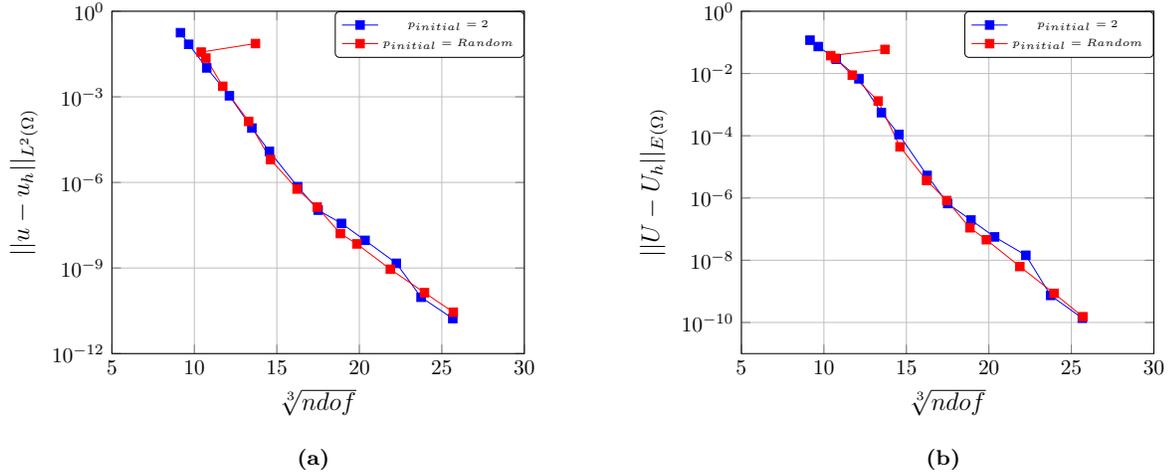

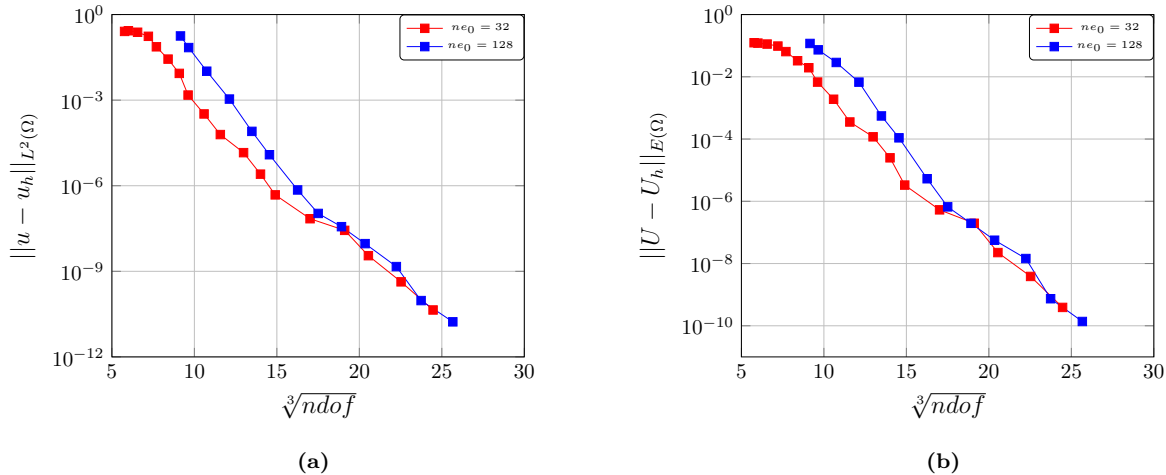
\begin{figure}[h!]
\begin{subfigure}[b]{0.5\textwidth}
\begin{tikzpicture}[scale=0.8]
		\begin{semilogyaxis}[xmin=5,xmax=30, ymin=1e-12,ymax=1,xlabel=\large{$\sqrt[3]{ndof}$},ylabel=\large{$||u-u_h||_{L^{2}(\Omega)}$},grid=major,legend style={at={(1,1)},anchor=north east,font=\tiny,rounded corners=2pt}]

          \addplot [color =red,mark=square*] table[x= ndof, y=err_l2, col sep = comma] {Data_hp/Boundary_layer⁩/exp/L2_error_BL_MNS_0p5_Hpnew.txt};
          	
           \addplot [color =blue,mark=square*] table[x= ndof, y=err_l2, col sep = comma] {Data_hp/Boundary_layer⁩/exp/L2_error_patchsolve_hp_1to20_DLS_128el_BL.txt};		
           
		\legend{$ne_{0} = 32$,$ne_{0} = 128$}
		\end{semilogyaxis}
	\end{tikzpicture}
	\caption{}
\end{subfigure}
\begin{subfigure}[b]{0.5\textwidth}
\begin{tikzpicture}[scale=0.8]
		\begin{semilogyaxis}[xmin=5,xmax=30, ymin=1e-11,ymax=1,xlabel=\large{$\sqrt[3]{ndof}$},ylabel=\large{$||U-U_h||_{E(\Omega)}$},grid=major,legend style={at={(1,1)},anchor=north east,font=\tiny,rounded corners=2pt}]
		
\addplot [color =red,mark=square*] table[x= ndof, y=EE, col sep = comma] {Data_hp/Boundary_layer⁩/exp/EE_error_BL_MNS_0p5_Hpnew.txt};

\addplot [color =blue,mark=square*] table[x= ndof, y=EE, col sep = comma] {Data_hp/Boundary_layer⁩/exp/EE_error_patchsolve_hp_1to20_DLS_128el_BL.txt};
						
	\legend{$ne_{0} = 32$,$ne_{0} = 128$}
		\end{semilogyaxis}
\end{tikzpicture}
\caption{}
\end{subfigure}	
\caption{Convergence plots of (a) $L^2$ error in $u_h$ and (b) energy error using scaled V-norm with different initial mesh.} \label{convergence_BL_diffmesh_hp}
\end{figure}

\Cref{fig:contourAndPdist} shows a contour plot of the solution obtained on  an adapted mesh  as well as the corresponding polynomial degree distribution. In~\cref{convergence_BL_scaled_math_norm_hp}, we present the convergence results comparing the $h-$adaptation algorithm~\cite{CHAKRABORTY20221} and proposed new $hp-$adaptation algorithm. On continuously increasing $\mathcal{N}_{h,p}$, $hp-$adaptation produces better convergence compared to the $h-$adaptation with constant polynomial orders $p = 1-5$. For all convergence curves, we plot error against $\sqrt[3]{ndof}$ to verify exponential convergence (${\Vert e \Vert}_{X} \approx C e^{-bN^{1/3}}$, \cite{Guo1986} ). On increasing $\mathcal{N}_{h,p}$, initially the adaptation is dominated by $h-$refinement. Consequently, once the boundary layer is resolved, $p-$refinement starts taking place in the boundary layers along with $h-$refinements. The algorithm prefers higher-order polynomials in elements in the boundary layer, whereas it prescribes $p = 2$ away from the boundary layer. The nature of the analytical solution is approximately quadratic away from the boundary layer, at least up to machine precision, i.e., $u(x) \approx xy$ \cite{Ar2021}. Since with the assumption of shape-regular elements one can show the equivalence of the scaled V-norm with the $L^2$ norm of the field variables~\cite{Demkowicz2011a}, we expect to obtain the polynomial distribution reflecting the local behaviour of the solution variables. 

%Previously, while showing the convergence behavior, we kept increasing the $\mathcal{N}_{h,p}$ by $30\%$. 

In the next numerical experiment, we perform the adaptation while keeping $\mathcal{N}_{h,p}$ constant. Here, we have limited $p_{max}$ (maximum polynomial order in the $hp$ mesh) at $p_{max}=10$. Since $u \in C^{\infty}$, $p_{max}$ will otherwise increase indefinitely. Through this numerical experiment, we intend to observe the interplay between $h-$ and $p-$refinements, which is measured by computing the average polynomial order  ($p_{avg}$) in the mesh. The evolution of $p_{avg}$ with subsequent adaptations is shown in~\cref{hp_fixed_dof_ne_pavg} for different values of $\epsilon$. For $\epsilon = 0.005$, $ h-$refinements dominate initial adaptations, and it is only after some initial $h-$refinement in the boundary layers that $p-$refinements start dominating along with $h-$refinements. On increasing $\epsilon$, the algorithm performs simultaneous $p-$adaptation and $h-$adaptation as the boundary layers are smooth enough so as to not require increased spatial resolution for $p-$adaptations. This interplay between two adaptive processes is visible from the gradual decrease in the number of elements and subsequent increase in $p_{avg}$ (see~\cref{hp_fixed_dof_ne_pavg}). 

In many adaptation strategies, the initial conditions such as initial mesh or distribution of the polynomial order may affect the performance. In the proposed $hp-$adaptation algorithm, arbitrary coarsening and refinements can be done due to re-meshing. This reduces the dependency of the current mesh on previous iterations. To demonstrate this, we present two instances of variations in initial conditions. First, we use different initial meshes with the same initial polynomial order of approximation (see~\cref{convergence_BL_diffmesh_hp}). In second instance, the same initial mesh is used, but one mesh with a random distribution of initial order of approximation (see~\cref{convergence_BL_randompolyorder}). We observe no deviation in the asymptotic behavior in the convergence plots, showing the proposed methodology's robustness towards these perturbations.

\subsection{Goal oriented adaptation: Gaussian peak}
\begin{figure}[H]
\begin{center}
\includegraphics[scale=0.25]{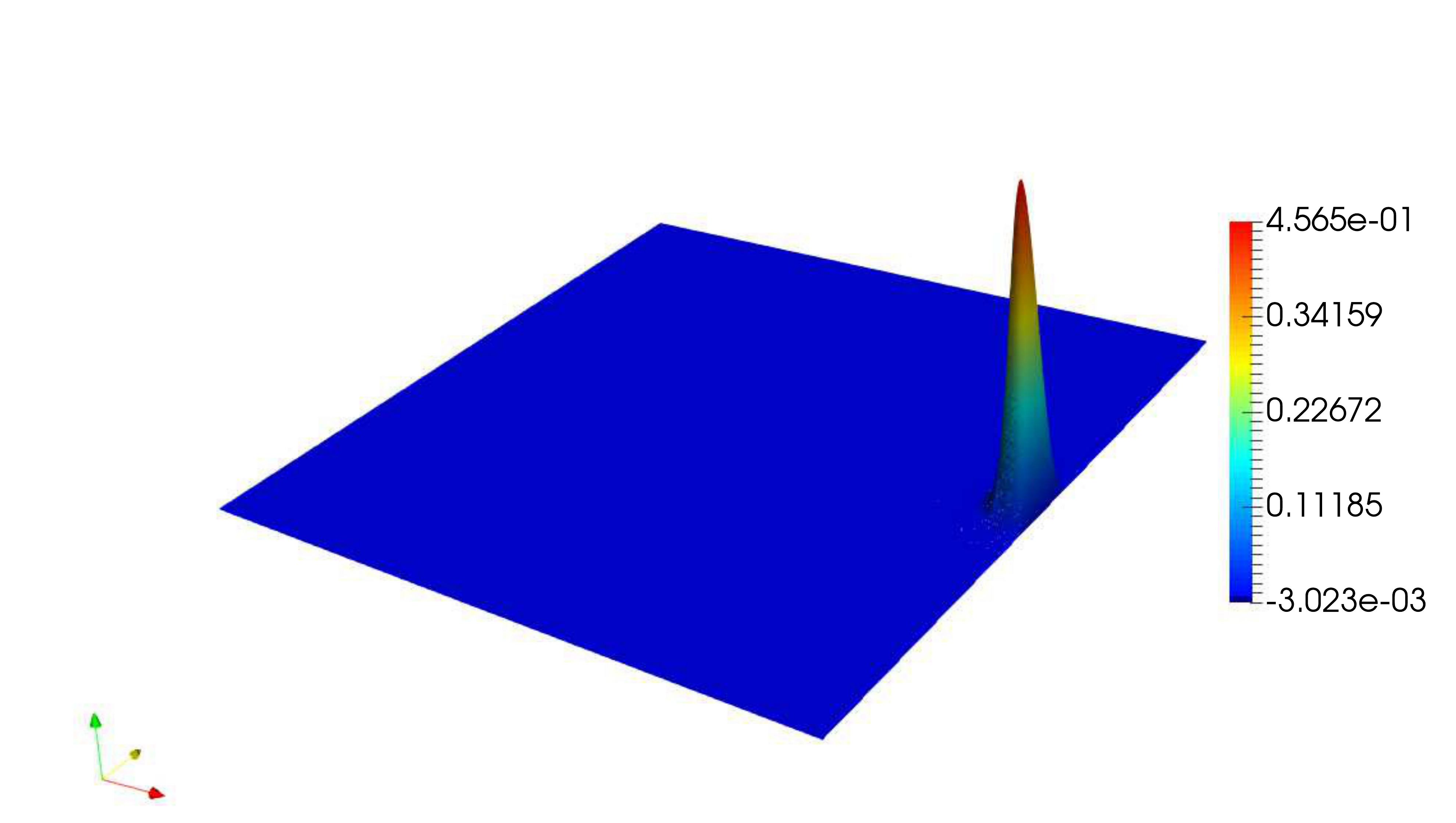}
\end{center}
\caption{Contour showing $j_{\Omega}(\mathbf{x})$.} \label{goalgaussian}
\end{figure}

Next, we present results showing the performance of the proposed $hp-$adaptation algorithm for goal-oriented adaptation. The primal problem is the same as that of~\cref{blhp}. We consider the solution-dependent (target) functional 
\begin{equation}
J(u) = \int_\Omega j_{\Omega}(\mathbf{x}) u(\mathbf{x}) \, d\mathbf{x}, \label{volumetargethp}
\end{equation}
where
\begin{equation}
j_{\Omega}(\mathbf{x}) = e^{-\alpha \left({(x - x_c)}^2 + {(y - y_c)}^2\right)}.
\end{equation}
The target is thus given by a weighted volume integral, where the weight is a Gaussian peak centered at $(x_c,y_c) = (0.99,0.5)$. We choose $\alpha = 1000$, leading to a strong localization of the peak (see~\cref{goalgaussian}). This leads to the dual problem 
\begin{align}
-\beta \cdot {\nabla} \bar{\eta}-\epsilon{\nabla}^2 \bar{\eta} &= j_{\Omega}\qquad && \mathrm{in}\ \Omega = {(0,1)}^2,  \label{adjstrngeqhp}\\
\bar{\eta} &= 0 && \mathrm{on}\ \partial \Omega,  \label{adjbndhp}
\end{align}
where $\beta = {[1, \,1]}^T$ and $\epsilon > 0$. 

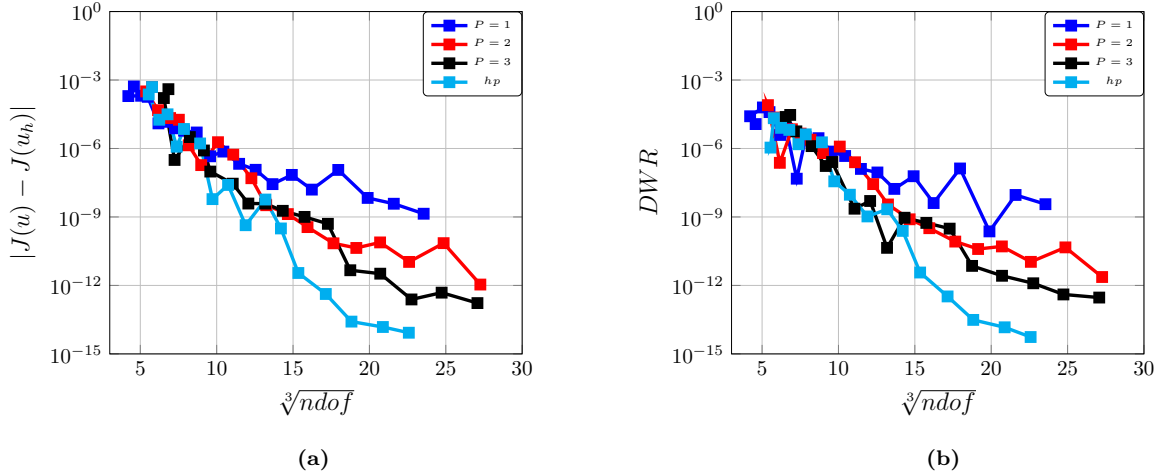
\begin{figure}[H]
	\begin{subfigure}[b]{0.5\textwidth}
		\begin{tikzpicture}[scale = 0.8]
			\begin{semilogyaxis}[xmin=3,xmax=30, ymin=1e-15,ymax=1,xlabel=\large{$\sqrt[3]{ndof}$},ylabel=\large{$\vert J(u) - J(u_h) \vert$},grid=major,legend style={at={(1,1)},anchor=north east,font=\tiny,rounded corners=2pt}]
				\addplot [color = blue,mark=square*,ultra thick] table[x=ndof, y=target_err,col sep = comma]{⁨Data_hp⁩/TargetBasedAdaptatioHp⁩/expconv/target_error_s_p_plus_1_gaussianpeak_eps_0p0005_alpha_1000_p1_exp.txt⁩};
				\addplot [color = red,mark=square*,ultra thick] table[x=ndof, y=target_err,col sep = comma]{⁨Data_hp⁩/TargetBasedAdaptatioHp⁩/expconv/target_error_s_p_plus_1_gaussianpeak_eps_0p0005_alpha_1000_p2_exp.txt};
				\addplot [color = black,mark=square*,ultra thick] table[x=ndof, y=target_err,col sep = comma]{⁨Data_hp⁩/TargetBasedAdaptatioHp/expconv/target_error_s_p_plus_1_gaussianpeak_eps_0p0005_alpha_1000_p3_exp.txt⁩};
				\addplot [color = cyan,mark=square*,ultra thick] table[x=ndof, y=target_err,col sep = comma]{⁨Data_hp⁩/TargetBasedAdaptatioHp⁩/expconv/target_error_gaussian_peak_hp.txt⁩};
				\legend{$P =1$,$P =2$,$P =3$,$hp$}
			\end{semilogyaxis}
		\end{tikzpicture}
		
		\caption{}\label{convergence_gaussian_peak_error_hp}
	\end{subfigure}
	\begin{subfigure}[b]{0.5\textwidth}
		\begin{tikzpicture}[scale = 0.8]
			\begin{semilogyaxis}[xmin=3,xmax=30, ymin=1e-15,ymax=1,xlabel=\large{$\sqrt[3]{ndof}$},ylabel=\large{$DWR$},grid=major,legend style={at={(1,1)},anchor=north east,font=\tiny,rounded corners=2pt}]       
				\addplot [color = blue,mark=square*,ultra thick] table[x=ndof, y=DWR_est,col sep = comma]{⁨Data_hp⁩/TargetBasedAdaptatioHp⁩/expconv/target_error_s_p_plus_1_gaussianpeak_eps_0p0005_alpha_1000_p1_exp.txt⁩};
				\addplot [color = red,mark=square*,ultra thick] table[x=ndof, y=DWR_est,col sep = comma]{⁨Data_hp⁩/TargetBasedAdaptatioHp⁩/expconv/target_error_s_p_plus_1_gaussianpeak_eps_0p0005_alpha_1000_p2_exp.txt};
				\addplot [color = black,mark=square*,ultra thick] table[x=ndof, y=DWR_est,col sep = comma]{⁨Data_hp⁩/TargetBasedAdaptatioHp⁩/expconv/target_error_s_p_plus_1_gaussianpeak_eps_0p0005_alpha_1000_p3_exp.txt⁩};
            					\addplot [color = cyan,mark=square*,ultra thick] table[x=ndof, y=DWR_est,col sep = comma]{⁨Data_hp⁩/TargetBasedAdaptatioHp⁩/expconv/target_error_gaussian_peak_hp.txt⁩};
				\legend{$P =1$,$P =2$,$P =3$,$hp$}
			\end{semilogyaxis}
		\end{tikzpicture}
		\caption{}\label{convergence_gaussian_peak_dwr_hp}
	\end{subfigure}
	\caption{Convergence plots for (a) error in target functional  and (b) dual weighted residual (DWR) using the scaled V-norm.} \label{convgaussianpeaksmoothing}
\end{figure}

\begin{figure}[h!]
\begin{center}
\includegraphics[scale=0.22]{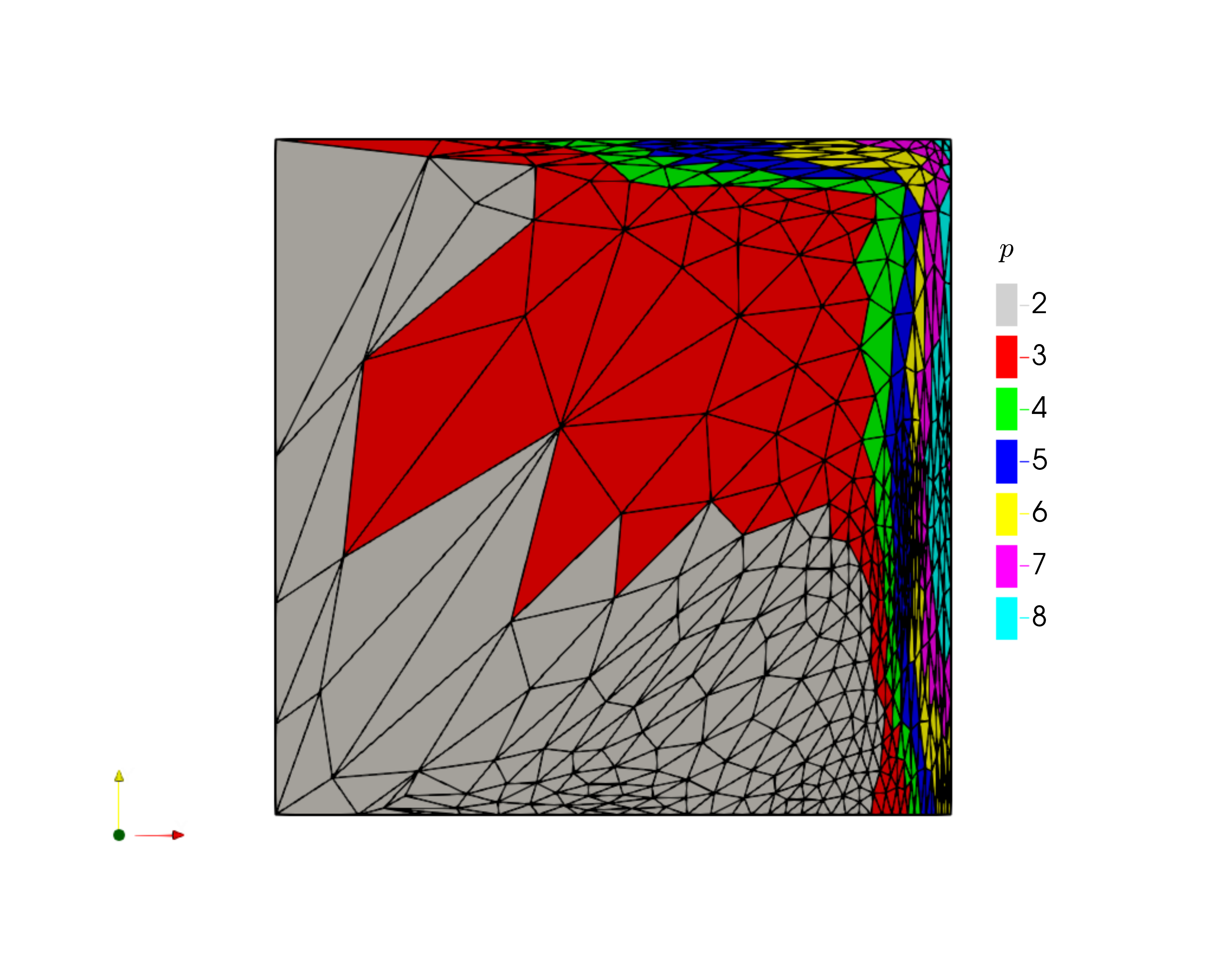}
\end{center}
\caption{Polynomial distribution on an adapted mesh with 11513 degrees of freedom.} \label{adaptedmeshgaussianpeak}
\end{figure}

In~\cref{convergence_gaussian_peak_error_hp} and~\cref{convergence_gaussian_peak_dwr_hp}, we present the convergence plots comparing $h-$adaptation \cite{CHAKRABORTY20221} to the proposed $hp-$adaptive method. We obtain exponential convergence for $hp-$adaptation, thus outperforming $h-$adaptation. 
%The presence of geometrically graded meshes can account for this superior performance.\gm{I'm not sure for smooth solutions like this you need, or obtain, geometrically graded meshes} 
We present a snapshot of one such mesh in~\cref{adaptedmeshgaussianpeak}. Note that both $h-$ and $p-$refinements mainly take place in the right boundary layer. As we previously observed in~\cref{blhp}, the $h-$refinements precede $p-$refinements, thus following a similar trajectory. The difference here is that only a portion of the boundary layer undergoes $hp-$refinement, which the algorithm deems necessary for resolving the target functional. We can notice the saturation of the $hp-$convergence curves near machine precision in~\cref{convergence_gaussian_peak_error_hp} and~\cref{convergence_gaussian_peak_dwr_hp}.

Next, we perform a numerical experiment where we keep $\mathcal{N}_{h,p}$ fixed and then run the adaptations. This numerical experiment aims to examine the algorithm's capacity to distribute the Dofs optimally. It takes only a few adaptations to reach nearly machine precision in terms of the error in target functional.
\begin{table}[h!]
\captionsetup{justification=centering,margin=2cm}
\centering
\begin{tabular}{|c|c|c|c|}
\hline
Adaptation & Ndof  & ${\vert J(u) - J(u_h) \vert}$ & $DWR$ \\ \hline
 0          & 3072 & 1.3599e-04                       & 2.55956e-05                       \\ \hline
2          & 2698 & 2.62413e-09                       & 1.16551e-09                        \\ \hline
4          & 3271 & 1.2955e-11                        & 1.69234e-11                       \\ \hline
6          & 3045 & 2.92165e-12                       & 2.71505e-12                       \\ \hline
8          & 3276 & 3.98067e-14                       & 1.31114e-13                        \\ \hline

\end{tabular}
\caption{Adaptation Vs. error for constant complexity using scaled V-norm ($\mathcal{N}_{h,p} = \int_{\Omega} w(\mathbf{x}) d(\mathbf{x}) \, d\mathbf{x} = 3072.0$).} \label{fixed_cost_target_a}
\end{table}

\subsection{Test case II: inverse tangent - flux target}
\begin{figure}[h!]
\begin{center}
\includegraphics[scale=0.22]{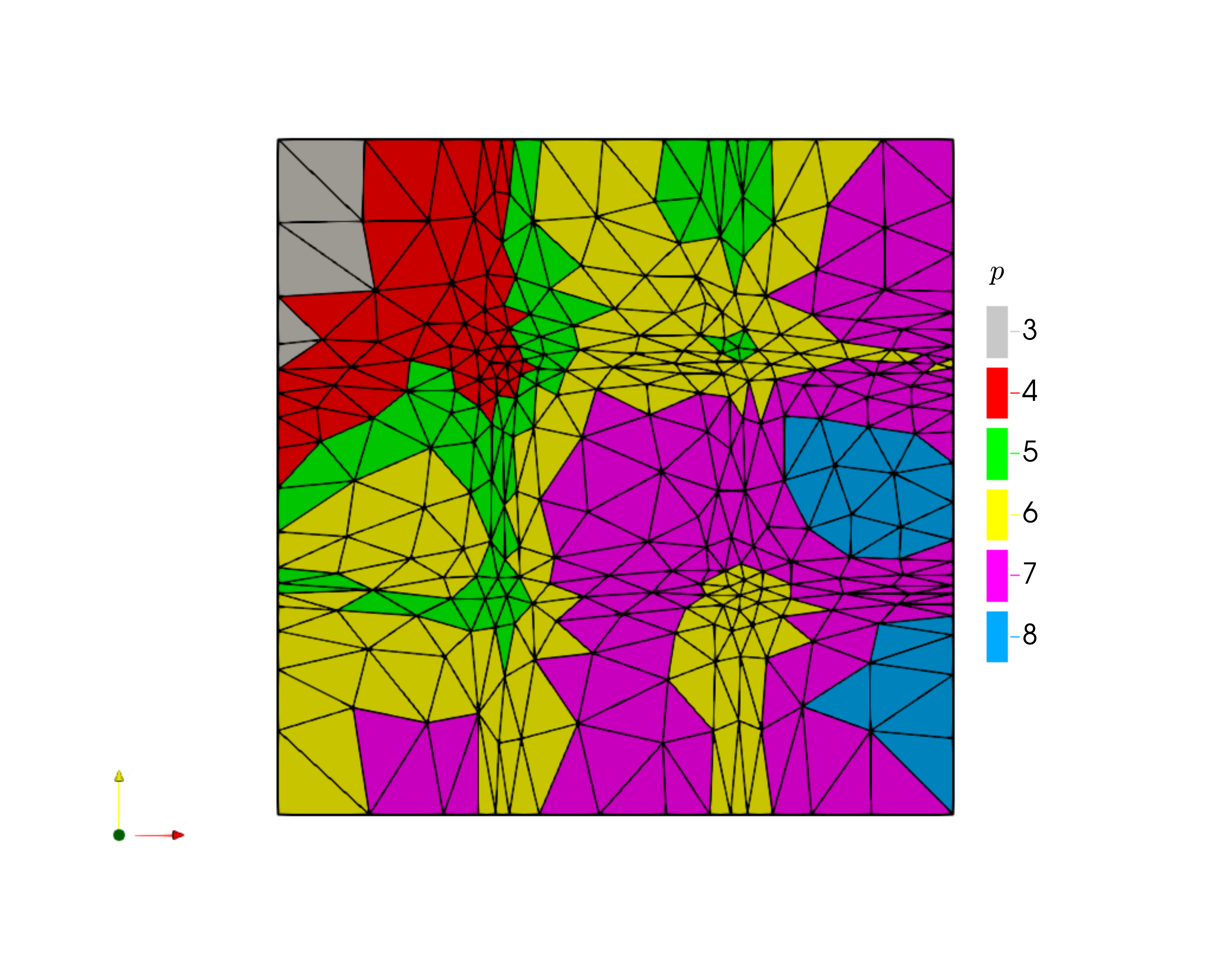}
\end{center}
\caption{Polynomial distribution on an adapted mesh with 15339 degrees of freedom.} \label{adaptedmeshfluxhp}
\end{figure}

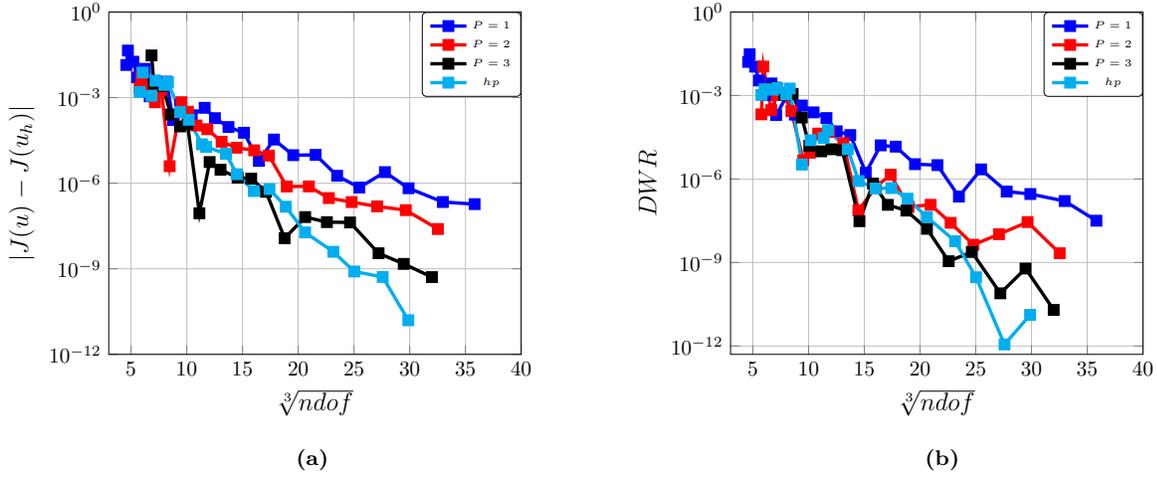
\begin{figure}[h!]
	\begin{subfigure}[b]{0.5\textwidth}
		\begin{tikzpicture}[scale = 0.8]
			\begin{semilogyaxis}[xmin=3,xmax=40, ymin=1e-12,ymax=1,xlabel=\large{$\sqrt[3]{ndof}$},ylabel=\large{$\vert J(u) - J(u_h) \vert$},grid=major,legend style={at={(1,1)},anchor=north east,font=\tiny,rounded corners=2pt}]
				\addplot [color = blue,mark=square*,ultra thick] table[x=ndof, y=target_err,col sep = comma]{⁨Data⁩_hp/TargetBasedAdaptatioHp⁩/expconv/targetbasedadaptation_s_p_plus_1_squarejump_Nadarajah_regularized_alpha_50_rightbndflux_eps_0p01_p1.txt⁩};
				\addplot [color = red,mark=square*,ultra thick] table[x=ndof, y=target_err,col sep = comma]{⁨Data⁩_hp/TargetBasedAdaptatioHp⁩/expconv/targetbasedadaptation_s_p_plus_1_squarejump_Nadarajah_regularized_alpha_50_rightbndflux_eps_0p01_p2.txt};
				\addplot [color = black,mark=square*,ultra thick] table[x=ndof, y=target_err,col sep = comma]{⁨Data⁩_hp/TargetBasedAdaptatioHp/expconv/targetbasedadaptation_s_p_plus_1_squarejump_Nadarajah_regularized_alpha_50_rightbndflux_eps_0p01_p3.txt⁩};
				\addplot [color = cyan,mark=square*,ultra thick] table[x=ndof, y=target_err,col sep = comma]{⁨Data⁩_hp/TargetBasedAdaptatioHp⁩/expconv/targetbasedadaptation_s_p_plus_1_squarejump_nadarajah_alpha50_hp.txt⁩};
				\legend{$P =1$,$P =2$,$P =3$,$hp$}
			\end{semilogyaxis}
		\end{tikzpicture}
		\caption{}\label{convergence_flux_error_hp}
	\end{subfigure}
	\begin{subfigure}[b]{0.5\textwidth}
		\begin{tikzpicture}[scale = 0.8]
			\begin{semilogyaxis}[xmin=3,xmax=40, ymin=0.5e-12,ymax=1,xlabel=\large{$\sqrt[3]{ndof}$},ylabel=\large{$DWR$},grid=major,legend style={at={(1,1)},anchor=north east,font=\tiny,rounded corners=2pt}]       
				\addplot [color = blue,mark=square*,ultra thick] table[x=ndof, y=DWR_est,col sep = comma]{⁨Data⁩_hp/TargetBasedAdaptatioHp⁩/expconv/targetbasedadaptation_s_p_plus_1_squarejump_Nadarajah_regularized_alpha_50_rightbndflux_eps_0p01_p1.txt⁩};
				\addplot [color = red,mark=square*,ultra thick] table[x=ndof, y=DWR_est,col sep = comma]{⁨Data⁩_hp/TargetBasedAdaptatioHp⁩/expconv/targetbasedadaptation_s_p_plus_1_squarejump_Nadarajah_regularized_alpha_50_rightbndflux_eps_0p01_p2.txt};
				\addplot [color = black,mark=square*,ultra thick] table[x=ndof, y=DWR_est,col sep = comma]{⁨Data⁩_hp/TargetBasedAdaptatioHp/expconv/targetbasedadaptation_s_p_plus_1_squarejump_Nadarajah_regularized_alpha_50_rightbndflux_eps_0p01_p3.txt⁩};
            		\addplot [color = cyan,mark=square*,ultra thick] table[x=ndof, y=DWR_est,col sep = comma]{⁨Data⁩_hp/TargetBasedAdaptatioHp⁩/expconv/targetbasedadaptation_s_p_plus_1_squarejump_nadarajah_alpha50_hp.txt⁩};
				\legend{$P =1$,$P =2$,$P =3$,$hp$}
			\end{semilogyaxis}
		\end{tikzpicture}
		\caption{}\label{convergence_flux_dwr_hp}
	\end{subfigure}
	\caption{Convergence plots for (a) error in target functional  and (b) dual weighted residual (DWR) using scaled V-norm.} \label{convnadarajahflux}
\end{figure}

This numerical experiment compares $h-$adaptation~\cite{CHAKRABORTY20221} and the proposed $hp-$adaptation algorithm for a target given by a boundary integral. The governing PDE is the same as in the boundary layer test case (see~\cref{blhp}). 
%We reiterate the PDE along with boundary conditions next.
%\begin{equation}
%\begin{aligned}
%\beta \cdot {\nabla}u-\epsilon{\nabla}^2u &= s(\mathbf{x}) \qquad&& \mathrm{in}\ \Omega = {(0,1)}^2, \\
%u &= g_D && \mathrm{on}\ \partial \Omega, 
%\end{aligned}
%\end{equation}
%where $\beta = {[1,1]}^T$. 
Here the source term $s(\mathbf{x})$ is chosen in such a way that the exact solution is given by
\begin{equation}
u(\mathbf{x}) = \left( tan^{-1}(\alpha(x - x_1)) + tan^{-1}(\alpha (x_2 - x)) \right)  \left( tan^{-1}(\alpha(y -y_1)) + tan^{-1}(\alpha (y_2 - y)) \right), \notag
\end{equation}
where $x_1 = y_1 = \frac{1.0}{3.0}$, $x_2 = y_2 = \frac{2.0}{3.0}$, $\alpha = 50.0 $ and $\epsilon = 0.01$. The boundary condition is obtained from the exact solution. The target functional is given by:
\begin{equation}
J({u}) = \int_{\partial \Omega} j_{\partial \Omega}(\mathbf{x}) {\nabla u} \cdot \mathbf{n} ds,
\end{equation}
where
\begin{equation}
j_{\partial \Omega} = \begin{cases}
1 \qquad \, \mathbf{n} = \left( 1,0 \right) \\
0 \qquad otherwise
\end{cases}.
\end{equation}
The dual problem has no volumetric source, and the weighting function $j_{\partial \Omega}(\mathbf{x})$ appears in the boundary condition for the dual problem:
\begin{align}
-\beta \cdot {\nabla} \bar{\eta}-\epsilon{\nabla}^2 \bar{\eta} &= 0 \qquad && \mathrm{in}\ \Omega = {(0,1)}^2,  \label{adjstrngeqbhp}\\
\bar{\eta} &= j_{\partial \Omega} && \mathrm{on}\ \partial \Omega.  \label{adjbndbhp}
\end{align}
In~\cref{adaptedmeshfluxhp}, we present the polynomial distribution of an $hp-$adapted mesh. Since we are adapting to resolve the flux at $x = 1.0$ and the convection is in direction ${[1,1]}^T$, the majority of the $hp-$refinement takes place in the right lower diagonal half of the domain (Since the flux on the right boundary depends upon the inlet conditions of the bottom edge at $y=0$, the dual error estimate gives more weighting to the elements in this half.). In~\cref{convergence_flux_error_hp} and~\cref{convergence_flux_dwr_hp}, we present the convergence results. Again we observe  exponential convergence in the case of $hp-$adaptations.

\subsection{Test case III: L-shaped domain}
Next, we present the classical L-shaped domain Poisson problem:
\begin{equation}
\begin{aligned}
-{\nabla}^2 u &= s\quad && \mathrm{in}\ \Omega = {(-1,1)}^2 \setminus [0,1] \times [-1,0], \\
u &= g_D && \mathrm{on} \ \partial \Omega.
\end{aligned}
\end{equation}
The source term $s(\mathbf{x})$ is chosen in such a way that the exact solution is given by:
\begin{equation}
u(\mathbf{x}) = r^{\frac{2}{3}}sin\left(\frac{2}{3}\theta\right) \quad where \quad \theta = tan^{-1}\left(\frac{y}{x}\right)  \quad \text{and} \quad r = \sqrt{x^2 + y^2}.
\end{equation}

The boundary conditions are taken from the exact solution. This test case serves to demonstrate the performance of $hp-$adaptivity in the presence of a singularity, where the additional flexibility of varying the local order of approximation leads to dramatically improved results compared to $h-$only adaptation. In the current example, when using uniform refinement, it is expected to achieve a convergence of $O(h^{\frac{4}{3}})$ \cite{crsing} in the $L^2$ sense in $u$.  In \cite{Yano2012}, it has already been shown that higher-order convergence i.e. $p+1$ (for $h-$adaptation) can be achieved using exponentially graded meshes. % However, the question arises whether one can do better than achieving a $p+1$ order of convergence?

In \cite{Guo1986} and \cite{DaVeiga2018}, it has been shown that exponential convergence can be achieved with a sequence of geometrically refined $hp-$meshes in the presence of a corner singularity. However, these meshes are typically hand-crafted and require a priori information (polynomial and size distribution), contrary to an automatic mesh adaptation procedure. Thus, generating such meshes with optimal gradation in size and polynomial distribution is a challenging problem for any automatic mesh adaptation algorithm.

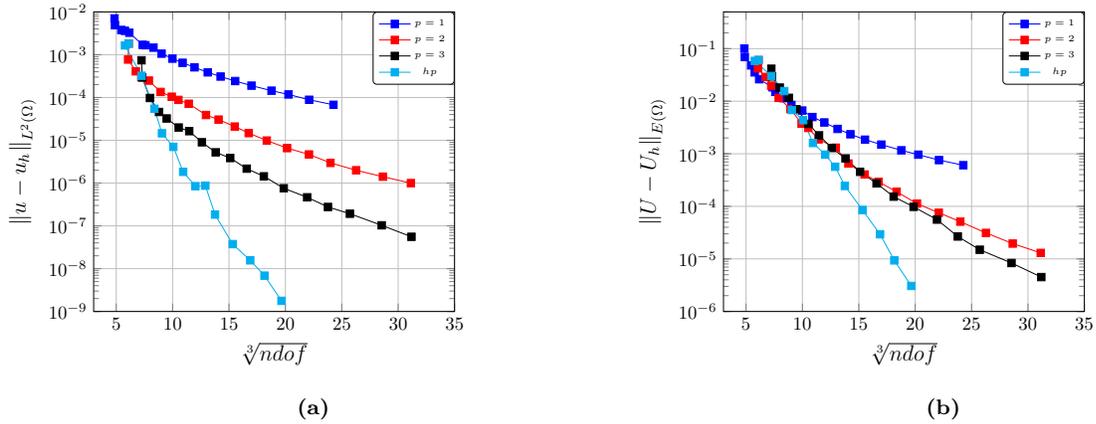
\begin{figure}[H]
\begin{subfigure}[b]{0.5\textwidth}
\begin{tikzpicture}[scale=0.7]
		\begin{semilogyaxis}[xmin=3,xmax=35, ymin=1e-9,ymax=0.01,xlabel=\large{$\sqrt[3]{ndof}$},ylabel=\large{${ \Vert u-u_h \Vert}_{L^{2}(\Omega)}$},grid=major,legend style={at={(1,1)},anchor=north east,font=\tiny,rounded corners=2pt}]
		\addplot[color = blue,mark=square*] table[x= ndof, y=err_l2, col sep = comma] {Data_hp/L_shaped⁩/expconv/L2_error_Lshaped_Domain_MNS_0p5_p1_hp.txt};
		\addplot [color = red,mark=square*] table[x= ndof, y=err_l2, col sep = comma] {Data_hp/L_shaped⁩/expconv/L2_error_Lshaped_Domain_MNS_0p5_p2_hp.txt};
		\addplot [color = black,mark=square*] table[x= ndof, y=err_l2, col sep = comma] {Data_hp/L_shaped⁩/expconv/L2_error_Lshaped_Domain_MNS_0p5_p3_hp.txt};
\addplot [color = cyan,mark=square*] table[x= ndof, y=err_l2, col sep = comma] {Data_hp/L_shaped⁩/expconv/L2_error_Lshaped_MNS_0p5_Hpnew_1to20.txt};

%		\legend{$p = 1$,$p = 2$,$p = 3$,$Hp_{\beta = 2.0}$,$Hp_{\beta = 2.5}$}
			\legend{$p = 1$,$p = 2$,$p = 3$,$hp$}
		\end{semilogyaxis}
	\end{tikzpicture}
	\caption{}
\end{subfigure}
\begin{subfigure}[b]{0.5\textwidth}
\begin{tikzpicture}[scale=0.7]
		\begin{semilogyaxis}[xmin=3,xmax=35, ymin=1e-6,ymax=0.5,xlabel=\large{$\sqrt[3]{ndof}$},ylabel=\large{${ \Vert U-U_h \Vert}_{E(\Omega)}$},grid=major,legend style={at={(1,1)},anchor=north east,font=\tiny,rounded corners=2pt}]
		\addplot[color = blue,mark=square*]  table[x= ndof, y=EE, col sep = comma] {Data_hp/L_shaped⁩/expconv/EE_error_Lshaped_Domain_MNS_0p5_p1_hp.txt};
		\addplot [color = red,mark=square*] table[x= ndof, y=EE, col sep = comma]  {Data_hp/L_shaped⁩/expconv/EE_error_Lshaped_Domain_MNS_0p5_p2_hp.txt};
			\addplot [color = black,mark=square*] table[x= ndof, y=EE, col sep = comma]  {Data_hp/L_shaped/expconv⁩/EE_error_Lshaped_Domain_MNS_0p5_p3_hp.txt};
\addplot [color =cyan,mark=square*] table[x= ndof, y=EE, col sep = comma]  {Data_hp/L_shaped/expconv/EE_error_Lshaped_MNS_0p5_Hpnew_1to20.txt};

%		\legend{$p = 1$,$p = 2$,$p = 3$,$Hp_{\beta = 2.0}$,$Hp_{\beta = 2.5}$}
			\legend{$p = 1$,$p = 2$,$p = 3$,$hp$}
		\end{semilogyaxis}
\end{tikzpicture}
\caption{}
\end{subfigure}	
\caption{Convergence plots of (a) $L^2$ error in $u_h$ and (b) energy error using scaled V-norm.} \label{convergence_scaledvnorm_Lshaped_plota}
\end{figure}

\begin{figure}[h]
\begin{subfigure}[b]{0.33\textwidth}
\begin{tikzpicture}[scale=0.6]
		\begin{semilogyaxis}[xmin=3,xmax=35, ymin=1e-6,ymax=0.1,xlabel=\large{$\sqrt[3]{ndof}$},ylabel=\large{$ {\Vert u-u_h \Vert}_{L^{\infty}(\Omega)}$},grid=major,legend style={at={(1,1)},anchor=north east,font=\normalsize,rounded corners=2pt}]
		\addplot[color = blue,mark=square*] table[x= ndof, y=Linf,col sep = comma] {Data_hp/L_shaped⁩/expconv/L2_inf_error_Lshapeddomain_p1.txt};
		\addplot [color = red,mark=square*] table[x= ndof, y=Linf, col sep = comma] {Data_hp/L_shaped⁩/expconv/L2_inf_error_Lshapeddomain_p2.txt};
		\addplot [color = black,mark=square*] table[x= ndof, y=Linf, col sep = comma] {Data_hp/L_shaped⁩/expconv/L2_inf_error_Lshapeddomain_p3.txt};			
		\addplot [color = cyan,mark=square*] table[x= ndof, y=Linf, col sep = comma] {Data_hp/L_shaped⁩/expconv/all_error_L_shaped_Hp.txt};

%		\legend{$p = 1$,$p = 2$,$p = 3$,$Hp_{\beta = 2.0}$,$Hp_{\beta = 2.5}$}
	\legend{$p = 1$,$p = 2$,$p = 3$,$hp$}
		\end{semilogyaxis}
	\end{tikzpicture}
	\caption{}
\end{subfigure}
\begin{subfigure}[b]{0.33\textwidth}
\begin{tikzpicture}[scale=0.6]
		\begin{semilogyaxis}[xmin=3,xmax=35, ymin=5e-7,ymax=0.2,xlabel=\large{$\sqrt[3]{ndof}$},ylabel=\large{$ {\vert u-u_h \vert}_{H^{1}(\Omega)}$},grid=major,legend style={at={(1,1)},anchor=north east,font=\normalsize,rounded corners=2pt}]
		\addplot[color = blue,mark=square*] table[x= ndof, y=H1_Seminorm_error, col sep = comma] {Data_hp/L_shaped⁩/expconv/allerror_Lshaped_hp_p1.txt};
		\addplot [color = red,mark=square*] table[x= ndof, y=H1_Seminorm_error, col sep = comma] {Data_hp/L_shaped⁩/expconv/allerror_Lshaped_hp_p2.txt};
		\addplot [color = black,mark=square*] table[x= ndof, y=H1_Seminorm_error, col sep = comma] {Data_hp/L_shaped⁩/expconv/allerror_Lshaped_hp_p3.txt};
								\addplot [color = cyan,mark=square*] table[x= ndof, y=H1_Seminorm_error, col sep = comma] {Data_hp/L_shaped⁩/expconv/all_error_L_shaped_Hp.txt};

%		\legend{$p = 1$,$p = 2$,$p = 3$,$Hp_{\beta = 2.0}$,$Hp_{\beta = 2.5}$}
	\legend{$p = 1$,$p = 2$,$p = 3$,$hp$}
		\end{semilogyaxis}
	\end{tikzpicture}
	\caption{}
\end{subfigure}
\begin{subfigure}[b]{0.33\textwidth}
\begin{tikzpicture}[scale=0.6]
		\begin{semilogyaxis}[xmin=3,xmax=35, ymin=5e-7,ymax=0.2,xlabel=\large{$\sqrt[3]{ndof}$},ylabel=\large{$ {\Vert u-u_h \Vert}_{H^1(\Omega)}$},grid=major,legend style={at={(1,1)},anchor=north east,font=\normalsize,rounded corners=2pt}]
		\addplot[color = blue,mark=square*]  table[x= ndof, y=H1_error, col sep = comma] {Data_hp/L_shaped⁩/expconv/allerror_Lshaped_hp_p1.txt};
		\addplot [color = red,mark=square*] table[x= ndof, y=H1_error, col sep = comma]  {Data_hp/L_shaped⁩/expconv/allerror_Lshaped_hp_p2.txt};
			\addplot [color = black,mark=square*] table[x= ndof, y=H1_error, col sep = comma]  {Data_hp/L_shaped/expconv⁩/allerror_Lshaped_hp_p3.txt};
								\addplot [color = cyan,mark=square*] table[x= ndof, y=H1_error, col sep = comma]  {Data_hp/L_shaped/expconv⁩/all_error_L_shaped_Hp.txt};

%		\legend{$p = 1$,$p = 2$,$p = 3$,$Hp_{\beta = 2.0}$,$Hp_{\beta = 2.5}$}
	\legend{$p = 1$,$p = 2$,$p = 3$,$hp$}
		\end{semilogyaxis}
\end{tikzpicture}
\caption{}
\end{subfigure}	
\caption{Convergence plots of (a) $L^{\infty}$ error, (b) $H^1$ seminorm error and (c) $H^1$ error using scaled V-norm.} \label{convergence_scaledvnorm_Lshaped_plotb}
\end{figure}
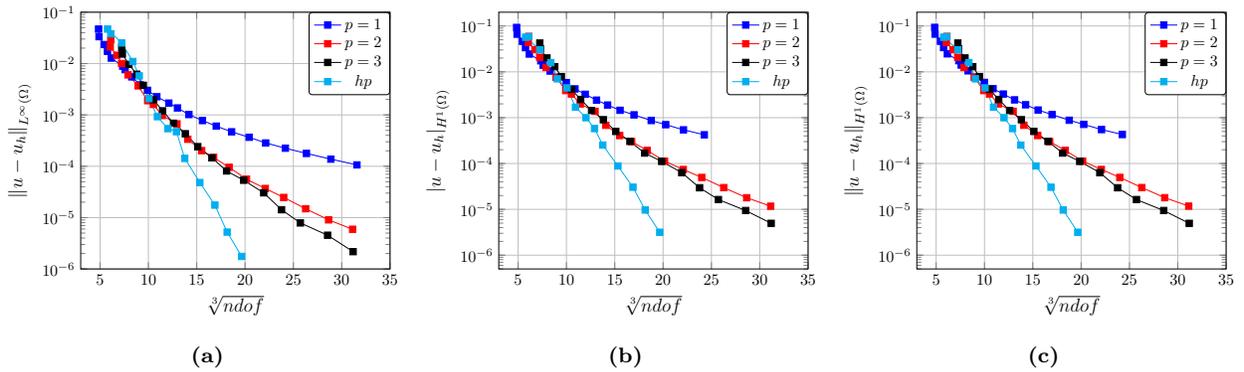

In ~\cref{convergence_scaledvnorm_Lshaped_plota} and ~\cref{convergence_scaledvnorm_Lshaped_plotb}, we present the convergence results. We observe exponential convergence in $L^2$ norm, energy norm, $L^{\infty}$, $H^1$ norm, and semi-norm. One can observe that $hp-$adaptation completely outperforms $h-$adaptation. 
%Convergence in $L^{\infty}$ becomes significant due to the presence of a singularity as the solution tends to blow up on approach towards the singularity. \gm{Note sure what this means}

\begin{figure}[H]
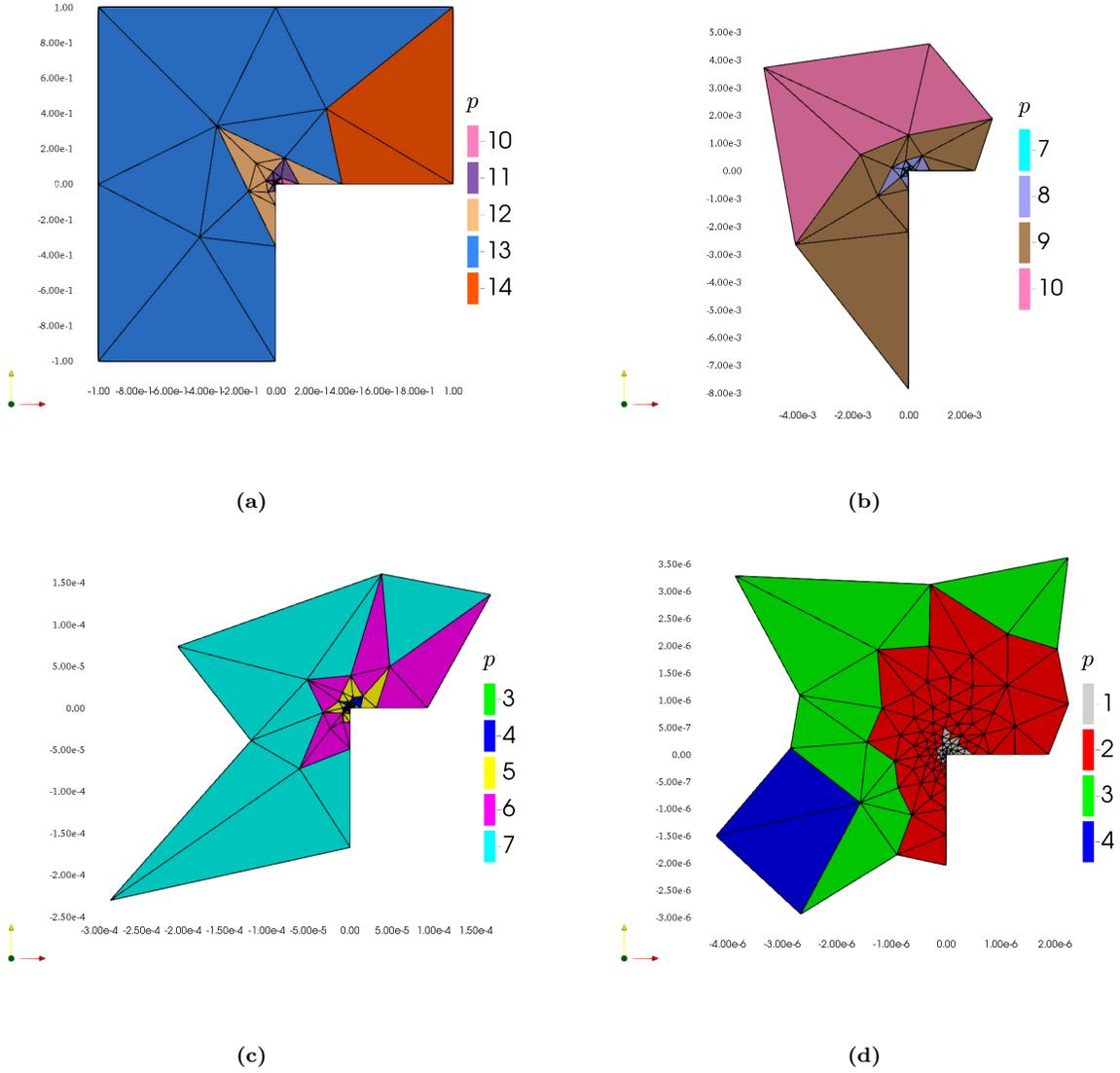

%\begin{center}
\begin{subfigure}[b]{0.5\textwidth}
\includegraphics[scale = 0.24]{Data_hp/L_shaped⁩/zoomedinview/level0-eps-converted-to.pdf}
\caption{}
\end{subfigure}	
\begin{subfigure}[b]{0.5\textwidth}
\includegraphics[scale = 0.24]{Data_hp/L_shaped⁩/zoomedinview/level1-eps-converted-to.pdf}
\caption{}
\end{subfigure}	
\begin{subfigure}[b]{0.5\textwidth}
\includegraphics[scale = 0.24]{Data_hp/L_shaped⁩/zoomedinview/level2-eps-converted-to.pdf}
\caption{}
\end{subfigure}	
\begin{subfigure}[b]{0.5\textwidth}
\includegraphics[scale = 0.24]{Data_hp/L_shaped⁩/zoomedinview/level3-eps-converted-to.pdf}
\caption{}
\end{subfigure}	

%\end{center}
\caption{Zoomed-in view showing polynomial order with (a) Level $0$, (b) Level $1$, (c) Level $2$ and (d) Level $3$ magnification  (with Level 0 denoting the least magnification and Level 3 denoting the most magnification).} \label{Lshapeddomain_polydist}
\end{figure}

In ~\cref{Lshapeddomain_polydist}, we present the polynomial distribution on an adapted mesh. Near the singularity, we observe the lowest order of approximation. As we move away from the singularity, the algorithm chooses higher polynomial orders due to the smoothness of primal variables. 
%This demonstrates the capability of the proposed methodology to correctly choose the local order of approximation.
%\begin{figure}[H]
%%\begin{center}
%\begin{subfigure}[b]{0.5\textwidth}
%\includegraphics[scale = 0.22]{Data_hp/L_shaped⁩/Lshapeddomain_polynomialdist_level3_247ne_7482dofs.eps}
%\caption{}
%\end{subfigure}	
%\begin{subfigure}[b]{0.5\textwidth}
%\includegraphics[scale = 0.22]{Data_hp/L_shaped⁩/Lshapeddomain_polynomialdist_level2_247ne_7482dofs.eps}
%\caption{}
%\end{subfigure}	
%\begin{subfigure}[b]{0.5\textwidth}
%\includegraphics[scale = 0.22]{Data_hp/L_shaped⁩/Lshapeddomain_polynomialdist_level1_247ne_7482dofs.eps}
%\caption{}
%\end{subfigure}	
%\begin{subfigure}[b]{0.5\textwidth}
%\includegraphics[scale = 0.22]{Data_hp/L_shaped⁩/Lshapeddomain_polynomialdist_level0_247ne_7482dofs.eps}
%\caption{}
%\end{subfigure}	
%
%%\end{center}
%\caption{Zoomed-in view showing polynomial order with (a) Level $0$ (b) Level $1$ (c) Level $2$ (d) Level $3$ magnification.} \label{Lshapeddomain_polydist}
%\end{figure}

Finally, we perform a numerical experiment where we keep the required DOFs fixed and adapt the mesh. This numerical experiment aims to demonstrate the reduction in error at a fixed cost. In ~\cref{Constant_DOF_Lshaped}, it can be observed that there is a reduction in both the $L^2$ error in $u$ and the energy error by three orders of magnitude.

\begin{table}[h]
\begin{center}
\begin{tabular}{|c|c|c|c|c|c|}
\hline
Adap no: & 0 &  3 & 5 & 7 & 10  \\ \hline
${\Vert u - u_h \Vert}_{L^2(\Omega)}$ & $0.000157763$ & $9.87e-07$ & $1.34e-07$& $1.13e-07$ & $1.89e-07$ \\ \hline
${\Vert U - U_h \Vert}_{E(\Omega)}$ & $0.0251805$ & $8.05e-04$ & $1.24e-04$& $3.66e-05$ & $1.98e-05$ \\ \hline
$p_{avg}$ & $2$ & $2.7$ & $4.01$& $5.57$ & $6.81$ \\ \hline
$\text{Ndof}$ & $3072$ & $2848$ & $3185$& $3238$ & $3308$ \\ \hline
\end{tabular}
\end{center}
\caption{Adaptation vs. error for constant complexity using scaled V-norm ($\mathcal{N}_{h,p} = \int_{\Omega} w(\mathbf{x}) d(\mathbf{x}) \, d\mathbf{x} = 3072.0$).} \label{Constant_DOF_Lshaped}
\end{table}

\section{Conclusion and outlook}
In this article, we present a continuous $hp-$mesh model that utilizes the inbuilt residual-based error estimator of DPG finite element schemes with optimal test functions. The model can drive both solution and target-based mesh adaptations. Moreover, the model can be extended to other minimal-residual finite element methods as long as one can localize the error estimate and obtain a polynomial representation of the error estimate. For non-linear problems, the boundary conditions needed for solving the local problems (see~\cref{polyselec}) on a patch need rigorous analysis. It will depend upon the nature of the non-linear problem and the linearization employed. An extension of the proposed methodology to generate optimal meshes for the compressible Navier-stokes equations is currently being investigated and will be presented in future work.

\section{Acknowledgment} 
Funding: This work was funded by the Deutsche Forschungsgemeinschaft (DFG, German Research Foundation) – 333849990/GRK2379 (IRTG Modern Inverse Problems).
%Here are two sample references: \cite{Feynman1963118,Dirac1953888}.

Declaration of interest: none

\section*{References}

\bibliography{refs}

\end{document}